\documentclass[preprint,prb,aps]{revtex4}
\usepackage{amsmath,amssymb}
\usepackage{graphicx}
\usepackage{dcolumn}
\usepackage{bm}
\begin{document}
\title{On the classical and quantum dynamics of a class of nonpolynomial oscillators}
	\author{V. Chithiika Ruby and M. Lakshmanan}
	\address{Department of Nonlinear Dynamics, School of Physics,
		Bharathidasan University, Tiruchirapalli - 620 024, India.}
	
\begin{abstract}
We consider two one dimensional nonlinear oscillators, namely (i) Higgs oscillator  and (ii) a $k$-dependent nonpolynomial rational potential, where $k$ is the constant curvature of a Riemannian manifold. Both the systems are of position dependent mass form, ${\displaystyle m(x) = \frac{1}{(1 + k x^2)^2}}$, belonging to the quadratic  Li$\acute{e}$nard type nonlinear oscillators. They admit different kinds of motions at the classical level. While solving the quantum versions of the systems, we consider a generalized position dependent mass Hamiltonian in which the ordering parameters of the mass term are treated as arbitrary. We observe that the quantum version of the Higgs oscillator is exactly solvable under appropriate restrictions of the ordering parameters, while the second nonlinear system is shown to be quasi exactly solvable using the Bethe ansatz method in which the arbitrariness of ordering parameters also plays an important role to obtain quasi-polynomial solutions. We extend the study to three dimensional generalizations of these nonlinear oscillators and obtain the exact solutions for the classical and quantum versions of the three dimensional Higgs oscillator. The three dimensional generalization of the quantum counterpart of the $k$-dependent nonpolynomial potential is found out to be quasi exactly solvable. 
\end{abstract}

\maketitle

\section{Introduction}
The classical nonlinear oscillator introduced by Mathews and Lakshmanan and characterized by the Lagrangian, 
\begin{equation}
L = \frac{1}{2}\frac{(\dot{x}^2 - \omega^2_0 x^2)}{(1 + \lambda x^2)}, 
\label{ml}
\end{equation}
where $\omega_0$ and $\lambda$ are parameters, has drawn considerable interest in the literature due to its unique simple harmonic oscillatory behavior \cite{mathews1974unique, mathews1975quantum}. The nonlinear system (\ref{ml}) and its generalizations are  continuously studied for many aspects such as $3$-dimensional \cite{lakshmanan1975quantum} and $d$-dimensional generalizations \cite{ranada2002harmonic, vkc2013, quesne2016}, rational extensions of the potentials  \cite{quesne2015, quesne17} and its inverted version \cite{Schulze2015}.  Higgs geometrically  interpreted the above nonlinear system which emerged while generalizing the Cartesian coordinates of Euclidean geometry. He also obtained a non-polynomial Lagrangian while defining the harmonic oscillator in  generalized coordinates which  arises on the gnomonic projection onto the tangent plane from the center of the sphere \cite{higgs1979dynamical}. 
The one-dimensional version of the Higgs oscillator is described by the Lagrangian, 
\begin{equation}
L = \frac{\dot{x}^2}{2(1 + k x^2)^2} - \frac{\omega_0}{2} x^2, 
\label{higg1d}
\end{equation} 
where $k$ is a parameter related to the curvature.  The Higgs oscillator has also attracted considerable attention since its introduction to the literature, mainly from the point of view of the notion of superintegrability. The two dimensional Higgs oscillator is considered to be an example of a superintegrable system in a non-flat space \cite{Bonatas}. We may  point out a few such studies on different integrable and superintegrable generalizations of Higgs oscillator \cite{ball1, ball2, ball3}, cuboctahedric Higgs oscillator \cite{aball2009}, quantum dynamics on its relativistic generalization \cite{Mohammadi} and also on the aspect of quantum exact solvability \cite{carinena2004non,carinena2007quantum, carinena2017},  construction of nonlinear coherent states with the observation of deformed oscillator algebra \cite{nlcs} and also on revealing generators of hidden symmetries and the conformal algebra \cite{OEvnin}. 

Both the above nonlinear systems correspond to specific examples for the  dynamics of systems in curved space. 
Interestingly, Cari\~{n}ena et al. obtained the two nonlinear systems, while studying the dynamics of a harmonic oscillator on the two dimensional Riemann spaces ($M^2_k$) of constant curvature (that is, sphere $S^2$ and hyperbolic plane ${\cal H}^2$). In this unified approach, the curvature $k$ was treated as a parameter and the corresponding Lagrangian of the harmonic oscillator potential in terms of  geodesic polar coordinates $(R, \Phi)$ on $M^2_k$ is expressed as \cite{carinena2004non,carinena2007quantum}
\begin{eqnarray}
{\cal L}(k) = \frac{1}{2}\left(v^2_R + S^2_k(R) v^2_{\Phi}\right) - \frac{1}{2}\omega^2_0 T^2_k(R), \label{carl}
\end{eqnarray}
where ${\displaystyle T_k = \frac{S_k}{C_k}}$, $S_k$ and $C_k$ are $k$-dependent trigonometric functions. They are defined as 
\begin{eqnarray}
C_k &=& \left \{
  \begin{tabular}{cc}
  $\cos(\sqrt{k}\;x),$ & $k>0,$  \\
  1, & $k = 0,$ \\
  $\cosh(\sqrt{-k}\;x),$ & $k<0,$ 
  \end{tabular}
\right.\\
S_k &=& \left \{
  \begin{tabular}{cc}
  $\frac{1}{\sqrt{k}}\sin(\sqrt{k}\;x),$ & $k>0,$  \\
  $x$, & $k = 0,$ \\
  $\frac{1}{\sqrt{-k}}\sinh(\sqrt{-k}\;x),$ & $k<0$. 
  \end{tabular}
\right.
\end{eqnarray}
Here $k>0$ denotes the spherical plane, whereas  $k < 0$ denotes hyperbolic plane and $k = 0$ denotes the Euclidean plane. 

(1) On the transformation $(R,\Phi) \rightarrow (r', \phi)$ as $r' = T_k(R)$, $\Phi = \phi$, the Lagrangian ${\cal L}(k)$ becomes
\begin{eqnarray}
{\cal L}_{H}(k) = \frac{1}{2}\left(\frac{v^2_{r'}}{(1 + k r'^2)^2} + \frac{v^2_{\phi}}{(1 + k r'^2)}\right)- \frac{1}{2}\omega^2_0 r'^2, 
\label{higgs2}
\end{eqnarray}
which is known as the two dimensional Higgs oscillator \cite{higgs1979dynamical}. 

(2) Similarly, on the transformation $(R,\Phi) \rightarrow (r, \phi)$ as $r = S_k(R)$, $\Phi = \phi$ with $k = - \lambda$, the Lagrangian ${\cal L}(k)$ becomes
\begin{eqnarray}
{\cal L}_{M}(\lambda) = \frac{1}{2}\left(\frac{v^2_{r}}{(1 + \lambda r^2)^2} + r^2 v^2_{\phi}\right)-
\frac{\omega^2_0}{2} \frac{r^2}{(1 + \lambda r^2)}, 
\label{higgs2}
\end{eqnarray}
which is known as the two dimensional Mathews-Lakshmanan oscillator \cite{mathews1974unique}. 

Classically, the kinetic energy term of the Mathews-Lakshmanan oscillator (MLO) depicts the dynamics of  a nonlinear $SU(2) X SU(2)$- chiral invariant Lagrangian expressed in Gasiorowicz-Geffen co-ordinates 
\cite{delbourgo1969infinities}, 
\begin{equation}
{\cal L} = \frac{1}{2}\left[(\partial_{\mu} {\bf \Phi}).(\partial_{\mu} {\bf \Phi}) + \frac{\lambda ( {\bf \Phi}.\partial_{\mu} {\bf \Phi})( {\bf \Phi}. \partial_{\mu} {\bf \Phi})}{(1 - \lambda {\Phi}^2)}\right], 
\label{gasiro}
\end{equation}
where $\lambda = constant$, and is expressed in flat space $\bf{\Phi} \rightarrow {\bf q}$. 

In another coordinate system, namely Schwinger coordinates, the same nonlinear chiral system (\ref{gasiro}) is expressed as  
\begin{equation}
\hspace{-1cm} \qquad{\cal L} = \frac{1}{2}\left[(\partial_{\mu} {\bf \Phi}).(\partial_{\mu}  {\bf \Phi}) - \frac{k}{(1 + k {\Phi}^2)}\left( {\Phi}^2 (\partial_{\mu}{\bf \Phi}). (\partial_{\mu} {\bf \Phi}) +\frac{ ({\bf \Phi}. \partial_{\mu} {\bf \Phi})({\bf \Phi}. \partial_{\mu}  {\bf \Phi})}{(1 + k {\Phi}^2)}\right)\right],   
\label{schwinger}
\end{equation}
where $k$ is constant \cite{delbourgo1969infinities}. 

In this work, we are interested to study the nonlinear systems associated with the Lagrangian (\ref{schwinger}), in the flat space, since it is related to the Higgs oscillator (\ref{higg1d}). In flat space, the Lagrangian takes the form 
\begin{equation}
L = \frac{1}{2}\left[\dot{\bf q}^2- \frac{k}{(1 + k {q}^2)}\left( {q}^2 {\bf {\dot q}}^2 +\frac{ ({\bf q}. \dot{\bf q})^2}{(1 + k{q}^2)}\right)\right].  
\label{schwingerq}
\end{equation}
We consider the one-dimensional version of the nonlinear chiral system (\ref{schwingerq}) corresponding to two different additional potentials, $V_{1,2}(x),$  
\begin{equation}
L = \frac{1}{2}\left[\frac{\dot{x}^2}{(1 + k {x}^2)^2}\right] - V_{1,2}(x),  
\label{schwingerx}
\end{equation}
where 
\begin{eqnarray}
V_1(x) = \frac{\omega^2_0 x^2}{2}, \qquad \qquad V_2(x) = \frac{\omega^2_0 x^2}{2 (1 + k x^2)^2}.  
\end{eqnarray}
Here, the potential $V_1(x)$ corresponds to the Higgs oscillator, while $V_2(x)$ is a non-polynomial potential that includes merely the mass term in its potential as in the case of Mathews-Lakshmanan oscillator (\ref{ml}). We also note here that the above type of Lagrangians involving potentials $V_1$ and $V_2$ have also been studied as superintegrable systems defined on $N$-dimensional space of constant curvature \cite{ball1,ball2,ball3}. 
The conjugate momentum for these systems (\ref{schwingerx}) is 
\begin{equation}
p = \frac{\partial L}{\partial {\dot{x}}} = \frac{\dot{x}}{(1 + k x^2)^2}, \label{momentum1d}
\end{equation} 
and the corresponding Hamiltonian can be written as 
\begin{equation}
H = \frac{(1 + k x^2)^2}{2} p^2 + V_{1,\;2}(x).
\label{schwinger-ham}
\end{equation}
The above Hamiltonians are of the form of position dependent mass systems, ${\displaystyle H = \frac{p^2}{2 m(x)} + V(x)}$, where $m(x)$ is the position dependent mass term. 
Such position dependent mass systems arise while studying the nonlinear chiral models in non-flat space. These systems also arise in connection with a modification of the Heisenberg uncertainty principle which accounts for the presence of gravity \cite{Heisenberg}.

In this work, we intend to study both the classical and quantum solvability of these one-dimensional systems. 
Both the above systems with the potentials $V_1(x)$ and $V_2(x)$ are classically of quadratic Li\'{e}nard-type one-dimensional nonlinear oscillators. They can be exactly solved classically. Here, the mass of the systems is position dependent, ${\displaystyle m(x) = \frac{1}{(1 + k x^2)^2}}$, where $k$ is constant. We have first employed semiclassical treatment to analyze the quantum solvability of the nonlinear systems. Recently, we studied the problem of removal of ordering ambiguity in position dependent mass quantum systems characterized by a generalized position dependent mass Hamiltonian which generalizes a number of  Hermitian as well as non-Hermitian ordered forms of the Hamiltonian \cite{chithiikaremoval, karthiga}. Motivated by this study, while analyzing the quantum dynamics of the nonlinear systems endowed with one-dimensional potentials, we use a generalized $2 N$-parameter kinetic energy operator in which we keep the ordering parameters as arbitrary.  Interestingly the above two systems can also be generalized to higher dimensions as well. Particularly, we will consider the three dimensional generalizations of these systems, both classically and quantum mechanically.

The manuscript is structured into three major segments as the classical, semiclassical and quantum studies of the above two one-dimensional nonlinear systems and their corresponding three dimensional generalizations. In sec.  \ref{class}, we study the classical dynamics of the one dimensional systems.  The quantum solvability of these systems is examined through semi-classical technique, namely the modified Bohr-Sommerfeld quantization rule in Sec. \ref{semi}.  In Sec. \ref{pdm}, we use the form of general ordered position dependent mass Hamiltonian and analyze the solvability of the nonlinear systems. We further extend our study to the three dimensional generalization of the two nonlinear systems in sections \ref{qhigg3d} and \ref{nlo3d}. In Sec. \ref{conc}, we present a summary of the results on the two systems. 

\section{\label{class} Classical dynamics of the one dimensional nonlinear chiral models}
Even though the two nonlinear systems have the same kinetic energy term, the difference in their potentials will lead to different solutions. We first consider the Higgs oscillator and then the nonpolynomial nonlinear oscillator system. 

\subsection{One dimensional Higgs oscillator}
The Lagrangian of the Higgs oscillator is of the form,  
\begin{equation}
\hspace{4.5cm}L = \frac{\dot{x}^2}{2(1 + k x^2)^2} - \frac{\omega^2_0}{2} x^2, \nonumber \hspace{6cm} (\ref{higg1d})
\end{equation} 
where $k$ and $\omega_0$ are parameters. 

The classical dynamics of the system (\ref{higg1d}) is described by the equation of motion, 
\begin{equation}
\ddot{x} - \frac{2 k x }{(1 + k x^2)} \dot{x}^2 + \omega^2_0\;{(1 + k x^2)^2}\;x = 0. 
\label{eom1}
\end{equation}   
It is of the form of the quadratic Li\'{e}nard type nonlinear equation, 
\begin{equation}
y'' + f(y) y'^2 + g(y) = 0. \label{lie} 
\end{equation}
By using the multiplication factor, $\dot{x}$, Eq. (\ref{eom1}) is integrated  to be 
\begin{equation}
\frac{\dot{x}^2}{(1 + k x^2)^2}+ \omega^2_0 x^2 = \epsilon = \text{constant}. 
\label{xdx}
\end{equation}

On implementation of the simple transformation, 
\begin{equation}
y = \frac{x}{\sqrt{1 + k x^2}}, 
\label{yx}
\end{equation}
equation (\ref{xdx}) is reduced to the form 
\begin{equation}
\dot{y}^2 + (\omega^2_0 +  \epsilon k) y^2 = \epsilon. 
\label{ydy2}
\end{equation}
When $k > 0$, $y$ is real and bounded between $0$ and ${\displaystyle \frac{1}{\sqrt{k}}}$  for all values of $x$, that is $-\infty <  x < \infty$ whereas for $k < 0$, $y \in (0, \infty)$ in the region ${\displaystyle -\frac{1}{\sqrt{|k|}} <  x < \frac{1}{\sqrt{|k|}}}$. 

\begin{figure}[!ht]
\vspace{0.5cm}
\begin{center}
\includegraphics[width=0.5\linewidth]{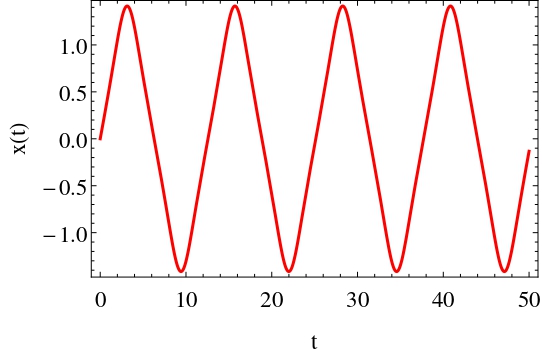}
\end{center}
\vspace{-0.3cm}
\caption{The plot of $x(t)$ obtained in (\ref{xt}) of the system (\ref{higg1d}) for $k= 0.5,\; A=1$ and $\Omega = 0.5$.} \label{first}
\vspace{-0.3cm}
\end{figure}

On integrating (\ref{ydy2}), we obtain  
\begin{equation}
\int \frac{d y }{\sqrt{1 - \frac{(\omega^2_0 +  \epsilon k)}{\epsilon}{y}^2}} = \sqrt{\epsilon}\; t + C, 
\label{int1}
\end{equation}
where $C$ is a constant. Consequently, we get 
\begin{equation}
y(t) = A \sin(\Omega t + C), 
\label{yt}
\end{equation}
where ${\displaystyle A^2 = \frac{\epsilon}{(\omega^2_0 +  \epsilon k)}}$ is the square of the amplitude and considered to be real and  ${\displaystyle \Omega^2 = \frac{\omega^2_0}{(1- k A^2)}}$ is the square of the frequency. The solution of  equation (\ref{eom1}) can be obtained from (\ref{yx}) and (\ref{yt}) as 
\begin{equation}
x(t) = \frac{A \sin(\Omega t + C)}{\sqrt{1 - k A^2 \sin^2(\Omega t + C)}},  
\label{xt}
\end{equation}
where ${\displaystyle |A|< \frac{1}{\sqrt{k}}}$ for $k>0$ and   $x(t)$ is periodic for all the values of $A$ when $k<0$. The explicit expression (\ref{xt}) of the one dimensional Higgs oscillator consisting of trigonometric function depicts not only the  solvability of the nonlinear system classically but also motivates us to study the exact solvability of its quantum version. The variable $x(t)$ is plotted in figure \ref{first}. 

\subsection{One dimensional nonlinear oscillator, $V_2(x)$}
Let us consider the Lagrangian of the potential, ${\displaystyle V_2(x) = \frac{\omega^2_0 x^2}{2\;(1 + k x^2)^2}}$, which is   
\begin{equation}
L = \frac{\dot{x}^2-\omega^2_0\;x^2}{2(1 + k x^2)^2}, 
\label{nlo1d}
\end{equation} 
where $k$ and $\omega_0$ are system parameters. The corresponding equation of motion is 
\begin{equation}
\ddot{x} - \frac{2 k x }{(1 + k x^2)} \dot{x}^2 + \omega^2_0 \frac{(1 - k x^2)\;x}{(1 + k x^2)} = 0, 
\label{eom2}
\end{equation}   
which is also of the form of the quadratic Li\'{e}nard type nonlinear equation, Eq. (\ref{lie}). Hence, the multiplication factor, ${\displaystyle \frac{\dot{x}}{(1+k x^2)^2}}$, reduces Eq. (\ref{eom2}) to the form 
\begin{equation}
\frac{\dot{x}^2}{(1 + k x^2)^2} + \frac{\omega^2_0 x^2}{(1 + k x^2)^2} = \epsilon = \text{constant}. 
\label{xdx2}
\end{equation}

To solve  Eq. (\ref{xdx2}), we use the transformation $\tau = \Omega t$ and $x(t) = A X(\tau)$, so that it can be re-expressed as 

\begin{equation}
\dot{X}^2(\tau) = \frac{\epsilon}{A^2\;\Omega^2} + \frac{(2\; k \epsilon - \omega^2_0)}{\Omega^2} X^2 + \frac{\epsilon\;k^2\;A^2}{\Omega^2} X^4,      
\label{xtau}
\end{equation}
 which results on integration to  
\begin{equation}
x(t) = A\;sn\left(\frac{\omega_0}{1 +  k A^2}\;t, m\right) 
\label{xtf}
\end{equation}
with 
\begin{eqnarray}
\epsilon = \frac{\omega^2_0\;A^2}{(1 + k A^2)^2}, \qquad \qquad m = k A^2.  
\end{eqnarray}

Using the identity \cite{ryzik}, ${\displaystyle sn(u, \; m) = \frac{(1 + {m_1}')\;sn(u_1,\;m_1)\;cn(u_1,\;m_1)}{dn(u_1,\;m_1)}}$, where $u = (1 + {m_1}')\;u_1$ and  ${\displaystyle m =\frac{1-{m_1}'}{1+{m_1}'}}$, and by substituting ${\displaystyle u_1 = \frac{\omega_0 t}{2}}$ and ${\displaystyle m'_1 = \frac{1-k A^2}{1+ k A^2}}$,  we can express the solution (\ref{xtf}) as 
\begin{equation}
x(t) = \frac{2\;A\;sn(\frac{\omega_0}{2}\;t, m_1) cn(\frac{\omega_0}{2}\;t, m_1)}{(1+k\;A^2) dn(\frac{\omega_0}{2}\;t, m_1)}, 
\label{xtfm}
\end{equation}
where the modulus ${\displaystyle m_1 = \frac{2\;\sqrt{k\;A^2}}{(1 + k \;A^2)}}$ lies  between  $0 \leq m_1 \leq 1$.  
 
\begin{figure}[!ht]
\vspace{0.5cm}
\begin{center}
\includegraphics[width=0.9\linewidth]{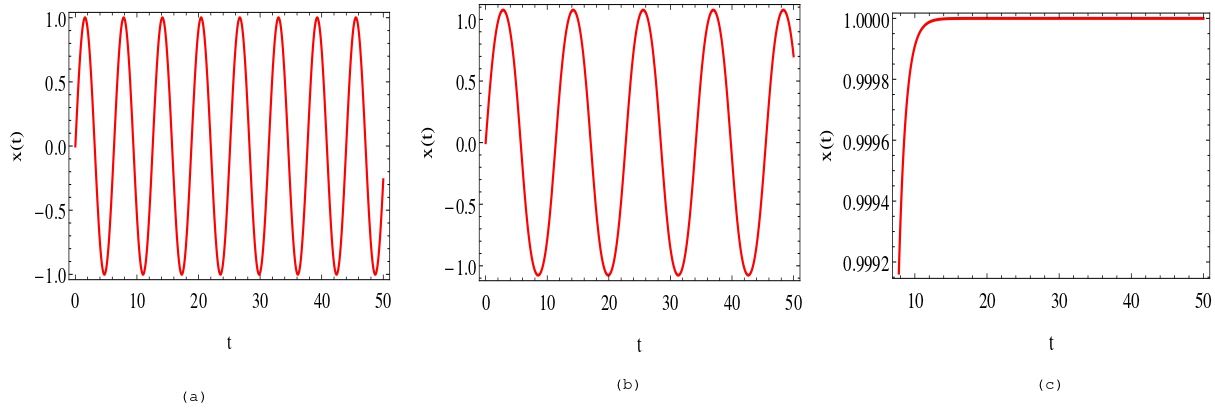}
\end{center}
\vspace{-0.3cm}
\caption{The plot of $x(t)$, (\ref{xtf}), of the system (\ref{nlo1d}) with different values of $k$ (a) $k =0$, (b) $k = 0.5$ and (c) $k = 1$ for $A = 1$ and $\omega_0 = 1$.} \label{second}
\vspace{-0.3cm}
\end{figure}

The solutions (\ref{xtf}) and (\ref{xtfm}) include the Jacobian elliptic functions $sn, \;cn$ and $dn$ which ensure us that  
the quantum counterpart of the system may be quasi exactly solvable. When $k \rightarrow 0$, the modulus $m = 0$ which reduces $x(t)$ (vide Eq. (\ref{xtf}) and (\ref{xtfm})) to be 
\begin{equation}
x(t) = A\;\sin(\omega_0 t), 
\label{xtf1}
\end{equation}
the solution of the linear harmonic oscillator as expected. The dynamics of the nonlinear system is dependent on $m_1$ which is depicted in  Figure. \ref{second}. 
 
\section{\label{semi}Semiclassical quantization}
The semiclassical study  often acts as a bridge between the study of classical dynamics and full quantum investigation of a system and often helps in the possibility of solving the quantum version of the system exactly. To quantize the system semiclassically, we use the modified Bohr-Sommerfeld quantization rule \cite{mbohr}, namely
\begin{equation}
\hspace{-2cm} \qquad \qquad  \qquad \oint p \;dx = \left(n + \frac{1}{2}\right) h, \qquad n = 0, 1, 2, 3, ... , 
\label{bohr}
\end{equation}
where $h$ is the Planck's constant and $n$ is any non-negative integer and the integration is carried out 
over a closed orbit in the $(x, p)$ space. 

\subsection{Higgs oscillator}
We can express the canonically conjugate momentum of the system (\ref{higg1d}) by using the solution (\ref{xt}) as 
\begin{eqnarray}
p = A \Omega \cos(\Omega t + C) \left(1 - k A^2 \sin(\Omega t + C\right)^{\frac{1}{2}}.
\label{mom1}
\end{eqnarray}
We first determine the energy of the system ${\displaystyle E =\frac{\epsilon}{2}}$ from (\ref{xdx}) or (\ref{ydy2}) using the solution (\ref{xt}) as  
\begin{eqnarray}
E = \frac{\omega^2_0\;A^2}{1 - k A^2}.
\label{enhigg}
\end{eqnarray}
Using $x(t)$, vide (\ref{xt}) in (\ref{momentum1d}), we can express the integral in (\ref{bohr}) to be  of the form, 
\begin{eqnarray}
I &=& \epsilon \oint \frac{dt}{(1 + (1+ k A^2) \tan^2(\Omega t + C))} \\
  &=& \frac{-\omega_0}{k}\left(1 - \frac{1}{\sqrt{1- k A^2}}\right),  
\end{eqnarray}
which on being substituted in (\ref{bohr}) yields 
\begin{eqnarray}
\frac{\omega^2_0\;A^2 }{(1 - k A^2)} =  2 \left(n + \frac{1}{2}\right) \hbar \omega_0 +\left(n^2 + n  + \frac{1}{4}\right)\hbar^2 k.
\label{A2}  
\end{eqnarray}
On substituting (\ref{A2}) in (\ref{enhigg}), we obtain the semiclassical energy spectrum of the Higgs oscillator (\ref{higg1d}) as 
\begin{eqnarray}
E_n =  \left(n + \frac{1}{2}\right) \hbar \omega_0 + \left(n^2 + n  + \frac{1}{4}\right) \frac{\hbar^2 k}{2}, \qquad n = 0, 1, 2, 3, ... .\label{higgs_en1d} 
\end{eqnarray}
Here the energy levels are quadratic as obtained by Higgs for the two dimensional case of the system (\ref{higg1d}) \cite{higgs1979dynamical}. In the limit $k \rightarrow 0$, the energy levels (\ref{higgs_en1d}) become the energy spectrum of the harmonic oscillator as the system (\ref{higg1d}) is reduced to that of the linear harmonic oscillator. 

\subsection{Semiclassical method for the second system (\ref{nlo1d})}
We now analyze the solvability of the system (\ref{nlo1d}) semiclassically. To do so, we use the momentum of the system (\ref{momentum1d}) as 
\begin{equation}
p = \frac{\partial L}{\partial \dot{x}} = \frac{\dot{x}}{(1 + k x^2)^2}. \nonumber
\end{equation}
By using the solution (\ref{xtf}), we can evaluate the integral in (\ref{bohr}) as 
\begin{eqnarray}
I &=& \frac{\omega^2_0 A^2}{(1 + k A^2)^2}\int^{4\;K(m_1)}_{0} cn^2(\omega_0 t, m_1) dt, \\
  &=& \frac{\omega_0}{4 k} (E(m_1) - K(m_1)),\label{i2}
\end{eqnarray}
where $y = cn(\omega_0 t, m_1)$ and $K(m_1)$ and $E(m_1)$ are complete elliptic integrals of the first kind and second kind, respectively. 
In terms of $m$, the relation (\ref{i2}) can be expressed as 
\begin{eqnarray}
\hspace{-1cm} \qquad I &=& \frac{\omega_0}{2 k}\left(\frac{E(m)}{(1+m)} - K(m)\right).\label{i2a}
\end{eqnarray}

Then,  $I$ can be related with quantum number $n$ through (\ref{bohr}) as 
\begin{equation}
\frac{\omega_0}{2 k}\left(\frac{E(m)}{(1+m)} - K(m)\right) = \left(n + \frac{1}{2}\right) h, \qquad n = 0, 1, 2, 3, ... .
\label{semihn}
\end{equation}
This relation (\ref{semihn}) implies that one has to solve for $m$ (and then $m_1$) and so for $A$ implicitly, which implies that only a numerical solution can be given but not an exact solution in general.
 
\section{\label{pdm}Quantum solvability of the nonlinear systems}
Most of the nonlinear oscillators defined in curved space are equivalently observed as position dependent mass systems on quantum transition.  They also include the quadratic Li\'{e}nard type nonlinear oscillators, for example, Mathews-Lakshmanan (ML) oscillator \cite{mathews1975quantum}, its generalizations \cite{midya}, its extensions \cite{barnana} and Higgs oscillator \cite{higgs1979dynamical}. 

The classical Hamiltonian $H(x, p)$ corresponding to (\ref{schwinger-ham}) is  
\begin{equation}
H = (1 + k x^2)^2 \frac{p^2}{2} + V(x), 
\label{non-ham}
\end{equation}
where  $p$ is the momentum (\ref{momentum1d}). 

In this section, we consider the two potentials, (1) ${\displaystyle V_1(x) = \frac{\omega^2_0 x^2}{2}}$, the Higgs oscillator and (2) the nonlinear system,${\displaystyle V_2(x) = \frac{\omega^2_0 x^2}{2 (1 + k x^2)^2}}$. The associated Hamiltonians are of the forms 
\begin{eqnarray}
H_1 &=& (1 + k x^2)^2 \frac{p^2}{2} + \frac{\omega^2_0 x^2}{2},  
\label{non-ham-higgs}
\end{eqnarray}
and 
\begin{eqnarray}
H_2 &=& (1 + k x^2)^2 \frac{p^2}{2} + \frac{\omega^2_0 x^2}{2 (1 + k x^2)^2},  
\label{non-ham-nlo}
\end{eqnarray}
respectively.

We observe that the Hamiltonian $H(x, p)$, specified by (\ref{non-ham}), is of position dependent mass type. 
To solve the system quantum mechanically, we consider the most general ordered form of the Hamiltonian operator that allows many numbers of possible mixtures of the fundamental 
term, $m^{\alpha} \hat{p} m^{\beta} \hat{p} m^{\gamma}$, where $\hat{p}$ is the momentum operator, $\hat{p} = -i\;\hbar\frac{d}{d x}$ defined in the Hilbert space $\mathbb{L}^2(\mathbb{R}, dx)$, and $x$ is the natural coordinate of the configuration space $\mathbb{R}$. It reads as  \cite{Trabelsi}
\begin{eqnarray}
\hat{H} = \frac{1}{2}\sum^N_{i = 1} w_i m^{\alpha_i} \hat{p} m^{\beta_i} \hat{p} m^{\gamma_i} + V(x), 
\label{geo1d}
\end{eqnarray}
where  $N$ is an arbitrary positive integer, and the ordering parameters $(\alpha_i, \beta_i, \gamma_i)$ should satisfy the constraint $\alpha_i +\beta_i +\gamma_i = -1,\;i =1, 2, 3, ... N$, and $w_i$'s  are real weights which are summed to be $1$. 

The above form globally connects all the Hermitian orderings and also provides a complete classification of Hermitian and non-Hermitian orderings \cite{Trabelsi}.  The kinetic energy operator of $\hat{H}$ in (\ref{geo1d}) possesses $2 N$ free ordering parameters.

The operator (\ref{geo1d}) is not Hermitian in general and can  be related to its Hermitian counterpart through a similarity transformation. The Hermitian Hamiltonian ${\hat{H}_{her}}$ is deduced  by performing the transformation 
\begin{equation}
\hat{ H}_{her}  = m^{\eta} \hat{H}_{non} m^{-\eta},  \label{hermitian} 
\end{equation}
with $2 \eta = \bar{\gamma} - \bar{\alpha}$, where $\bar{X}=(\bar{\alpha}, \bar{\beta}, \bar{\gamma})$ denotes the weighted mean value, ${\displaystyle \bar{X} = \sum_{i = 1}^N w_i X_i}$ and  $-\bar{\beta} = 1 + \bar{\alpha}+\bar{\gamma}$ 
(ref. \cite{chithiikaremoval}). The  Hermitian Hamiltonian  (\ref{hermitian}) for a potential $V(x)$ can be expressed as 
\begin{eqnarray}
\hspace{-1cm} \; \hat{H}_{her} = \frac{1}{2}\hat{p}\frac{1}{m}\hat{p} + \frac{\hbar^2}{2}\left[\left(\frac{\bar{\alpha}+\bar{\gamma}}{2}\right) \frac{d^2}{d x^2}\left(\frac{1}{m}\right) + \left(\overline{\alpha\gamma}+ \frac{1}{4} (\bar{\gamma} - \bar{\alpha})^2\right)  \left(\frac{d}{dx}\left(\frac{1}{m}\right)\right)^2 m\right]\nonumber \\ 
 \hspace{11cm}+ V(x).  
\label{geham_heo1d}
\end{eqnarray}
Since $[\hat{x}, \hat{p}] = i \hbar$, we assume the coordinate representation of  ${\displaystyle \hat{p} = - i \hbar \frac{d}{d x}}$. 

To solve the quantum system (\ref{non-ham}), we start with the time-independent Schr\"{o}dinger equation for the Hamiltonian (\ref{geham_heo1d}),  
\begin{eqnarray}
\hspace{-1cm} \psi{''} - \frac{m{'}}{m} \psi{'} + 
\left(\left(\frac{\bar{\alpha}+\bar{\gamma}}{2}\right)  \frac{m{''}}{m}-\left(\overline{\alpha\gamma}+\bar{\gamma}+\bar{\alpha}+\frac{1}{4}(\bar{\gamma} - \bar{\alpha})^2\right)  \frac{m{'^2}}{m^2}\right) \psi + \frac{2 m}{\hbar^2}\left(E -V(x)\right)\psi  = 0, \nonumber \\
\label{seg_he}
\end{eqnarray}
where ${\displaystyle ' = \frac{d}{dx}}$ and obtain the solution  for arbitrary values of the ordering parameters ${\bar \alpha}$ and ${\bar \beta}$.

We also note here that in the literature, different kinds of quantization procedures using, for example, the instantaneous Galilean invariance and Lagrangian of a nonlinear oscillator with the application of the Killing vector fields and associated Noether momenta, etc., have also been followed to study the quantum dynamics of position dependent mass Hamiltonians, \cite{mathews1974unique, carinena2017, levy}. In particular, Cari\~{n}ena et al. proposed a method of quantizing the PDM system by defining the Hilbert space, $\mathbb{L}^2(\mathbb{R}, d\mu)$ characterized by the square integrable functions with respect to the measure $d\mu$ related with the PDM. They first defined the metric associated with the PDM and then found out the existence of Killing vector fields for the PDM geodesic motion and the associated Noether momenta, which is conserved in a free force field. It is shown that the negative gradient of the potential force field is related to the time derivative of the Noether momenta rather to that of the canonical momenta \cite{omustafa}. Hence, Cari\~{n}ena et al., first quantized the Noether momenta and then the corresponding PDM Hamiltonian which is Hermitian on space $\mathbb{L}^2(\mathbb{R}, d\mu)$. However, in our analysis, we will be  concerned with the quantization of the PDM systems  in the $\mathbb{L}^2(\mathbb{R}, dx)$ space only as mentioned above by taking into appropriate orderings.

\subsection{Higgs oscillator, $H_1$}
For Higgs oscillator, the Hamiltonian is of the form,
\begin{equation}
\hspace{3cm}H = (1 + k x^2)^2 \frac{p^2}{2} + \frac{\omega^2_0\; x^2}{2}.  \nonumber \hspace{7cm}
(\ref{non-ham-higgs})
\end{equation}
The generalized time-independent Schr\"{o}dinger equation (\ref{seg_he}) for the Hamiltonian (\ref{non-ham-higgs}) becomes
\begin{eqnarray}
\hspace{-1cm} \; \psi{''} + \frac{4 k x}{(1+kx^2)} \psi{'} + 
\left(\frac{2 \eta_1 k -\frac{\omega^2_0 }{\hbar^2 k}}{(1 + k  x^2)}+ \frac{\frac{2 E}{\hbar^2}+ 4 \eta_2 k + \frac{\omega^2_0}{\hbar^2 k}}{(1 + k x^2)^2} \right)\psi  = 0, \qquad  \left(' = \frac{d}{dx}\right), 
\label{seg_he1}
\end{eqnarray}
where 
\begin{eqnarray}
\eta_1 &=& 5(\bar{\alpha} + \bar{\gamma}) - 8 \left(\bar{\alpha\gamma} + \bar{\alpha} + \bar{\gamma}+\frac{1}{4}(\bar{\gamma}- \bar{\alpha})^2\right),
\label{eta1}\\
\eta_2 &=&-3(\bar{\alpha} + \bar{\gamma}) + 4 
 \left(\bar{\alpha\gamma} + \bar{\alpha} + \bar{\gamma}+\frac{1}{4}(\bar{\gamma}- \bar{\alpha})^2\right).  
 \label{eta2}
\end{eqnarray}
We solve the equation (\ref{seg_he1}) for both the cases (i) $k < 0$ and (ii) $k > 0$. 

\subsection*{(i) Positive values of $k$}
On using the transformation, 
\begin{equation}
z = \frac{k\;x^2}{1+k\;x^2}, \qquad \psi(z) = (1-z)^{d}\;\phi(z), \label{solpsiz}
\end{equation}
we can express Eq. (\ref{seg_he1}) as 
\begin{eqnarray}
\hspace{-1cm} \; z(1 - z) \phi{''}(z) + \left( \frac{1}{2} - 2 d z\right) \phi{'}(z) +\left[\frac{\frac{\eta_1}{2}+d\left(d-\frac{3}{2}\right)-\frac{\mu^2}{4}}{1-z} - {\xi}\right] \phi(z) = 0, \; \left(' = \frac{d}{dz}\right), \nonumber \\
\label{hyperdz}
\end{eqnarray}
where
\begin{eqnarray}
{\xi} &=& \frac{E}{2\hbar^2 k } + \eta_2 +\frac{\mu^2}{4} - d(d-1), \label{xit}\\
\mu &=& \frac{\omega_0}{\hbar k}. \label{mu} 
\end{eqnarray}
With ${\displaystyle d = \frac{3}{4}\pm\frac{1}{2} \sqrt{\mu^2 -  2 \eta_1 + \frac{9}{4}}}$ and $\tilde{\mu} =  \sqrt{\mu^2 -  2 \eta_1 + \frac{9}{4}}$, we can reduce Eq. (\ref{hyperdz}) to the form
\begin{eqnarray}
\hspace{-1cm}z(1 - z) \phi{''}(z) + \left( \frac{1}{2} - \left(\pm\tilde{\mu} + \frac{3}{2}\right)z\right) \phi{'}(z) -\nu \left(-\nu\pm\tilde{\mu} + \frac{1}{2}\right)\phi(z) = 0, 
\label{hyperz}
\end{eqnarray}
where ${\xi}$ is now defined as,  
\begin{eqnarray}
{\xi} &=& -\nu \left(-\nu\pm\tilde{\mu}+\frac{1}{2}\right), \qquad \nu \in \mathbb{Z}. 
\label{energyd} 
\end{eqnarray}

Equation (\ref{hyperz}) is of the form of a hypergeometric equation, 
\begin{equation}
y(1- y)w''(y) + (c - (a + b +1)y)w'(y) - a b w(y) = 0, \qquad |y| < 1, \label{hyper}
\end{equation}
which admits the solution near $y = 0$ as
\begin{equation}
w(y) = C_1\;{}_2F_1(a, b; c; y) + C_2 y^{1-c}{}_2F_1\left(a-c+1, b-c+1; 2-c; y\right). 
\label{hyper-sol}
\end{equation}
On comparing (\ref{hyperdz}) and (\ref{hyperz}) we have 
\begin{equation}
a = \nu,\qquad b = -\nu\pm\tilde{\mu}+\frac{1}{2}, \qquad c = \frac{1}{2}. 
\end{equation}
 Hence the general solution of (\ref{hyperz}) can be written as   
\begin{equation}
\phi(z) =  C_1\;{}_2F_1\left(\nu, -\nu\pm\tilde{\mu}+\frac{1}{2}; \frac{1}{2}; z\right) + C_2 z^{1/2}{}_2F_1\left(\nu+\frac{1}{2},-\nu \pm \tilde{\mu}+1; \frac{3}{2}; z\right), \qquad |z|<1. 
\label{soln1}
\end{equation}
Here, $C_1$ and $C_2$ are arbitrary constants. 

On using  (\ref{soln1}) in (\ref{solpsiz}), we can express the solution for (\ref{seg_he1}) in terms of $z$ as  
\begin{eqnarray}
\hspace{-1cm} \; \psi(z) = (1-z)^{\frac{3}{4}\pm\frac{\tilde{\mu}}{2}} \left[C_1\;{}_2F_1\left(\nu, -\nu \pm \tilde{\mu}+\frac{1}{2}; \frac{1}{2}; z\right) + C_2 z^{1/2}{}_2F_1\left(\nu+\frac{1}{2},-\nu\pm\tilde{\mu}+1; \frac{3}{2}; z\right)\right]. \nonumber \\
\label{sol-siz}
\end{eqnarray}

To get bound state solutions, we apply the boundary condition $\psi(z) = 0$ at $z=1$ on (\ref{sol-siz}) which implies  that $d$ should be positive, that is, $\frac{3}{4}+\frac{\tilde{\mu}}{2}$. 
To obtain polynomial solution, we use the identity \cite{ryzik},
\begin{equation}
{\cal N} {}_2F_1\left(2 a, 2 b; a+b+\frac{1}{2}, \frac{1-\sqrt{z}}{2}\right) = C_1\;{}_2F_1\left(a, b; \frac{1}{2}; z\right) + C_2 z^{1/2}{}_2F_1\left(a+\frac{1}{2}, b+\frac{1}{2}; \frac{3}{2}; z\right),
\end{equation}
where $C_1$ and $C_2$ take the values
\begin{equation}
C_1 = \frac{\Gamma(a+b+\frac{1}{2})\sqrt{\pi}}{\Gamma(a+\frac{1}{2})\;\Gamma(b+\frac{1}{2})}\;{\cal N}, \qquad C_2 = \frac{\Gamma(a+b+\frac{1}{2})\sqrt{\pi}}{\Gamma(b)\;\Gamma(a)}\;{\cal N}. \label{constants}
\end{equation}
Here ${\cal N}$ is the normalization constant. Then, the solution (\ref{sol-siz}) can be expressed as 
\begin{equation}
\psi(z) = {\cal N}\;(1-z)^{\frac{3}{4}+\frac{\tilde{\mu}}{2}}\;{}_2F_1\left(2\nu, -2\nu+2\tilde{\mu}+1; \tilde{\mu}+1;\frac{1 - \sqrt{z}}{2}\right). 
\label{soln2a}
\end{equation}
The constants (\ref{constants}) are now of the forms, 
\begin{equation}
C_1 = \frac{\Gamma(\tilde{\mu}+1)\sqrt{\pi}}{\Gamma(\nu+\frac{1}{2})\;\Gamma(-\nu+\tilde{\mu}+1)}\;{\cal N}, \qquad C_2 = \frac{\Gamma(\tilde{\mu}+1)\sqrt{\pi}}{\Gamma(\nu)\;\Gamma(-\nu+\tilde{\mu}+\frac{1}{2})}\;{\cal N}. \label{constants-new}
\end{equation}

The auxiliary condition on the solution $\psi(z)$ that it should be square integrable requires $\psi(z)$  must be polynomial which can be achieved by choosing ${\displaystyle \nu = -\frac{n}{2}}$, where $n$ is a positive definite integer. Hence, Eq. (\ref{sol-siz}) becomes  
\begin{equation}
\psi_n(z) = {\cal N}_n\;(1-z)^{\frac{3}{4}+\frac{\tilde{\mu}}{2}}\;{}_2F_1\left(-n, n+2\tilde{\mu}+1; 1+\tilde{\mu}; \frac{1-\sqrt{z}}{2}\right), \quad n = 0, 1, 2, 3, .... \label{soln2}
\end{equation}
By using the transformation, \cite{ryzik}
\begin{equation}
{}_2F_1\left(c-a,c-b, c, y\right) = (1 - y)^{a+b-c}\;{}_2F_1\left(a, b;c; y\right),
\label{id1}
\end{equation}
we can express the solution (\ref{soln2}) as
\begin{equation}
\psi_n(z) = {\cal N}_n\;2^{\tilde{\mu}} (1 - z)^{3/4} \frac{(1 - \sqrt{z})^{\tilde{\mu}/2}}{(1+\sqrt{z})^{\tilde{\mu}/2}}\;{}_2F_1\left(-n-\tilde{\mu}, n+\tilde{\mu}+1;1+\tilde{\mu};\frac{1-\sqrt{z}}{2}\right).
\label{solnz}
\end{equation} 
Since the associated Legendre polynomial, $P^{-q}_{\lambda}(\sqrt{z})$, where $\lambda$ and $q$ are parameters, can be related to the hypergeometric function as
\begin{eqnarray}
\hspace{-1cm} \quad P^{-q}_{\lambda}(\sqrt{z}) =  \frac{1}{\Gamma(1 + {q})} \frac{(1 - \sqrt{z})^{q/2}}{(1 + \sqrt{z})^{q/2}}\;{}_2F_1\left(-\lambda,\lambda+1; 1+q; \frac{1-\sqrt{z}}{2}\right), \qquad q > 0, 
\label{solnx}
\end{eqnarray}
we can express the solution (\ref{solnz})  in the form
\begin{eqnarray}
\hspace{-1cm} \quad \psi_n(z) = {\cal N}_n\;2^{\tilde{\mu}}\Gamma{(1+\tilde{\mu})} (1 -z)^{\frac{3}{4}}\;P^{-\tilde{\mu}}_{n+\tilde{\mu}}(\sqrt{z}),  
\label{solnx}
\end{eqnarray}
with $\lambda = n + \tilde{\mu}$ and $q = \tilde{\mu}$. Then in terms of $x$, we have 
\begin{eqnarray}
\psi_n(x) = {\cal N}_n\;2^{\tilde{\mu}}\Gamma{(1+\tilde{\mu})} (1+ k\;x^2)^{-\frac{3}{4}}\;P^{-\tilde{\mu}}_{n+\tilde{\mu}}\left(\frac{\sqrt{k}\;x}{\sqrt{1+k\;x^2}}\right),\qquad k>0,  
\label{solnx}
\end{eqnarray}
where ${\displaystyle 0 < \frac{\sqrt{k}\;x}{\sqrt{1+k\;x^2}} < 1}$ provided that 
when $k>0,\;-\infty < x < \infty$. 
Here, the normalization constant ${\cal N}_n$  can be found out as 
\begin{eqnarray}
1 &=& \int^{\infty}_{-\infty} \psi^{*}_n(x) \psi_n(x) dx, \qquad k > 0 \\
{\cal N}_n &=& \left(\frac{\sqrt{k}\;\Gamma(n+2\tilde{\mu}+1)\;(2n+2\tilde{\mu}+1)}{n!\;2}\right)^{1/2}.
\end{eqnarray}
The corresponding energy eigenvalues $E_n$ can be obtained from (\ref{xit}) and (\ref{energyd}) with $\nu = -\frac{n}{2}$ as
\begin{eqnarray}
\frac{E}{2\hbar^2\;k} + \frac{\mu^2}{4} + \eta_2  - \left(\frac{\tilde{\mu}}{2} + \frac{3}{4}\right)\left(\frac{\tilde{\mu}}{2} - \frac{1}{4}\right) =-\nu \left(-\nu+\tilde{\mu}+\frac{1}{2}\right), 
\label{energyd1} 
\end{eqnarray}
which yields
\begin{equation}
E_n = \left(n + \frac{1}{2}\right) \hbar \sqrt{\omega^2_0 + \hbar^2 k^2 \left(\frac{9}{4}-2\eta_1\right)}+ \left(n^2 + n + 2\;\bar{\alpha} + 2\bar{\gamma} + \frac{3}{2}\right) \frac{\hbar^2 k}{2}, 
\label{energy-higgs1}
\end{equation}
where $\eta_1$ (vide Eq. (\ref{eta1})) is a function of the ordering parameters $\bar{\alpha}$ and $\bar{\gamma}$. Here the presence of ordering parameters $\bar{\alpha}$ and $\bar{\gamma}$ in the energy eigenvalues (\ref{energy-higgs1}) indicates that the different choices of ordering parameters results in different energy spectrum. Thus different choices of ordering parameters lead to different effective potentials of (\ref{non-ham-higgs}). Hence we have obtained a class of exactly solvable potentials  by using this procedure.  

In the limit $k \rightarrow 0$, we obtain the energy spectrum of the harmonic oscillator, $E_n = \left(n+\frac{1}{2}\right)\hbar \omega_0$, as the system (\ref{non-ham-higgs}) represents the simple harmonic oscillator in the same limit. 

In the recent work \cite{karthiga1}, the general ordered non-Hermitian ordered form (\ref{geo1d}) of the Hamiltonian corresponding to the Mathews-Lakshmanan oscillator (\ref{ml}) has been solved exactly.  The corresponding eigenfunctions are also  expressed  in terms of associated Legendre polynomials as   
\begin{equation}
\hspace{-2.5cm}\psi_n= \left\{\begin{array}{cc}
        N_n (1+ \lambda x^2)^{\frac{\bar{\gamma}-\bar{\alpha}}{2}}\; P_{n+\mu}^{-\mu}(\lambda^{\frac{1}{2}}x),\quad |x|<\lambda^{-\frac{1}{2}}\\
    \hspace{-1cm} 0, \quad |x|>\lambda^{-\frac{1}{2}},\,\, n =0, 1, 2, 3, ..., 
              \end{array}\right.\label{soln-ml} 
\end{equation}
where $\mu=\omega_0/\lambda \hbar$ and $N_n$ is the normalization constant. We observe that the discrete solutions of both the systems (vide (\ref{solnx}) and (\ref{soln-ml})) differ in their independent variables  which implies that the solutions of the Higgs oscillator can be related to that of the Mathews-Lakshmanan oscillator by an appropriate transformation. The corresponding energy spectrum of the the Mathews-Lakshmanan oscillator (\ref{ml}) is  
\begin{equation}
E_n = \left(n + \frac{1}{2}\right) \hbar \sqrt{\omega^2_0 + \hbar^2 \lambda^2 \left(4\bar{\alpha\;\gamma}+(\bar{\gamma}-\bar{\alpha})^2\right)}+ \left(n^2 + n -\;\bar{\alpha} -\bar{\gamma}\right) \frac{\hbar^2 \lambda}{2}. 
\label{legendre}
\end{equation}
The bound-state energy levels  of the Higgs oscillator are found to be a function of $k$  similar to the Mathews-Lakshmanan oscillator (\ref{ml}) in which the energy levels are also quadratic in $n$ and are  linearly dependent on the parameter $\lambda$. 

\subsection*{(ii) Negative values of $k$}
When $k < 0$, that is $k = -|k|$, the spatial region is divided into the following two regions: 
\begin{eqnarray}
\mbox{Region\;I}&:&\qquad -\frac{1}{\sqrt{|k|}} \leq  x \leq \frac{1}{\sqrt{|k|}}\\
\mbox{Region\;II}&:&\qquad  |x| > \frac{1}{\sqrt{|k|}}. 
\end{eqnarray}

We analyze the solvability of (\ref{seg_he1}) in accordance with the two regions. Eq. (\ref{seg_he1}) then  becomes    
\begin{eqnarray}
\hspace{-1cm}\psi{''} - \frac{4 |k| x}{(1-|k| x^2)} \psi{'} + 
\left(\frac{-2 \eta_1 |k| + \frac{\omega^2_0 }{\hbar^2 |k|}}{(1 -|k| x^2)}+ \frac{\frac{2 E}{\hbar^2} - 4 \eta_2 |k| - \frac{\omega^2_0}{\hbar^2 |k|}}{(1 -|k| x^2)^2} \right)\psi  = 0, 
\label{seg_hek-}
\end{eqnarray}
where $\eta_1$ and $\eta_2$ are defined in (\ref{eta1}) and (\ref{eta2}). 

\subsection*{(a)\;Region I: Bound States}
On using the transformation, 
\begin{equation}
z = |k|\;x^2, \qquad \psi(z) = (1-z)^{d}\;\phi(z), \label{solpsizk-}
\end{equation}
we can express the equation (\ref{seg_hek-}) as 
\begin{eqnarray}
\hspace{-1cm} \; z\;(1 - z) \phi{''}(z) + \left[\frac{1}{2} - \left(2d+\frac{5}{2}\right)z\right]\;\phi{'}(z) - \left( d -\frac{\tilde{\mu}}{2}+ \frac{3}{4}\right)\left(d+\frac{\tilde{\mu}}{2}+ \frac{3}{4}\right)\phi(z) = 0 
\label{eqk-}
\end{eqnarray}
with 
\begin{eqnarray}
\frac{E}{2\;\hbar^2 |k| } - \frac{\mu^2}{4} -  \eta_2+   d(d+1) &=& 0, \label{energyk-}\\
\hspace{5cm}\mu &=& \frac{\omega_0}{\hbar k}, \label{muk-} \qquad \mbox{and}\\  
\hspace{5cm}\tilde{\mu} &=& \sqrt{\mu^2 - 2 \eta_1 +\frac{9}{4}}. 
\end{eqnarray}

Equation (\ref{eqk-}) is of the form of the hypergeometric differential equation (\ref{hyper}) and hence the solution, $\psi(z)$ (vide (\ref{hyper-sol})), in terms of $x$ can be written as,
\begin{eqnarray}
\hspace{-1cm}\;\psi(x) = (1-|k|x^2)^d\left[ C_1\;{}_2F_1\left(d+\frac{3}{4}-\frac{\tilde{\mu}}{2},d+\frac{3}{4}+\frac{\tilde{\mu}}{2}; \frac{1}{2}; |k| x^2\right) \right. \nonumber \\
\hspace{2cm}\left. +C_2\;\sqrt{|k|}x\;{}_2F_1\left(d-\frac{1}{4}-\frac{\tilde{\mu}}{2}, d-\frac{1}{4}+\frac{\tilde{\mu}}{2};\frac{3}{2}; |k|x^2 \right)\right],
\label{psix1k-}
\end{eqnarray}
where $C_1$ and $C_2$ are parameters. 

We consider positive values of  $d$ to obtain bounded solutions.  And  the hypergeometric function of (\ref{psix1k-}) is reduced to  polynomial form so as to normalize the solution as 
\begin{eqnarray}
\hspace{-1cm} \qquad \psi(x) =  N \;2^{d-1}\frac{(1- \sqrt{|k|} x)^d}{(1+\sqrt{|k|}x)^{d+1}} {}_2F_1\left(\tilde{\mu}-\frac{1}{2}, \tilde{\mu}+\frac{1}{2}; 2d+2;\frac{1-\sqrt{|k|} x}{2}\right), \label{psik1}
\end{eqnarray}
where we have used the identities subsequently, 
\begin{eqnarray}
\hspace{-1cm} {}_2F_1\left(a, b; \frac{a+b+1}{2}; y\right) &=& \frac{\sqrt{\pi}\; \Gamma{\left(\frac{a+b+1}{2}\right)}}{\Gamma{\frac{a+1}{2}}\;\Gamma{\frac{b+1}{2}}}{}_2F_1\left(\frac{a}{2}, \frac{b}{2}; \frac{1}{2}; (2y -1)^2 \right) \nonumber \\
\hspace{-1cm} \; &+& (2 y -1) \frac{2\sqrt{\pi}\; \Gamma{\left(\frac{a+b+1}{2}\right)}}{\Gamma{\frac{a}{2}}\;\Gamma{\frac{b}{2}}}{}_2F_1\left(\frac{a+1}{2}, \frac{b+1}{2}; \frac{3}{2}; (2y -1)^2 \right), \label{identity1}\\
\hspace{-1cm} \; \hspace{1.6cm}{}_2F_1\left(a, b; c; y\right) &=& (1-y)^{c-a-b} {}_2F_1\left(c-a, c-b; c; y\right). \label{identity2}
\end{eqnarray}

Equation (\ref{psik1}) is reduced to be polynomial when $\tilde{\mu}+\frac{1}{2} -n=2d+2, $ $n = 0, 1, 2, ...$ is integer. Hence,  the value of  $d$ becomes ${\displaystyle d = \frac{1}{2}\left(-n+\tilde{\mu}-\frac{3}{2}\right)}$ to get bounded and normalizable eigenfunction. Now we use another identity, 
\begin{equation}
P^{m}_{\nu}(y) = \frac{1}{\Gamma{1-m}}\;\frac{(1+y)^{m/2}}{(1-y)^{m/2}}\;{}_2F_1\left(-\nu, \nu+1; 1-m; \frac{1-y}{2}\right)
\end{equation}
which transforms the solution (\ref{psik1}) to 
\begin{equation}
\psi_n(x) =  N_n \frac{\Gamma{\left(\frac{1}{2}-n+\tilde{\mu}\right)}}{\sqrt{(1-|k|\;x^2)}}\; P^{n-\tilde{\mu}+\frac{1}{2}}_{\tilde{\mu}-\frac{1}{2}}(\sqrt{|k|\;x)}, 
\label{psikn}
\end{equation}
where $P^{m}_{\nu}(y)$ is known to be the associated Legendre polynomial with $m$ and $\nu$ as parameters. Here, $n = 0, 1, 2, ...N$ and the upper limit $N < 2 \tilde{\mu}-1$ and the normalization constant $N_n$ is evaluated to be 
\begin{eqnarray}
N_n = \left(\frac{\sqrt{|k|}\;\left(n-\tilde{\mu}+\frac{1}{2}\right)\;\Gamma{(2\tilde{\mu}-n)}}{\Gamma{(n+1)}\;
2^{\frac{n}{4}-\frac{\tilde{\mu}}{4}+\frac{1}{8}}}\right)^{1/2},
\end{eqnarray}
since 
\begin{equation}
\int^{1}_{-1}\;\frac{P^{m}_{l}(y)\;P^{n}_{l}(y)}{\sqrt{1-y^2}}\;dy= \left\{\begin{array}{cc}
0, &  m \neq n, \\
\frac{(l+m)!}{m(l-m)!}, & m = n \neq 0, \\
\infty, & m = n =0. 
\end{array}\right. 
\end{equation}
Energy eigenvalues can be evaluated from (\ref{energyk-}) as  
\begin{equation}
E_n = \left(n + \frac{1}{2}\right)\hbar\omega_0\sqrt{1+ \frac{|k|^2 \hbar^2}{{\omega^2}_0}\;\left(\frac{9}{4}-2 \eta_1\right)} - \left(n^2 + n +2 \bar{\alpha} +  2 \bar{\gamma}+\frac{3}{2}\right)\frac{\hbar^2 |k|}{2}, \qquad n = 0, 1, 2, 3, ...N. 
\end{equation}
The energy eigenvalues $E_n$ are discrete and finite in number. And above the upper limit of $n$, that is $N$, the energy eigenvalue is continuous. 

\subsection*{(b) Region II: Continuous energy states}
We consider the region $|x| > \frac{1}{\sqrt{|k|}}$. Eq. (\ref{eqk-}) is in the form of hypergeometric equation (\ref{hyper}) which admits the solution near $z = \infty$ as 
\begin{eqnarray}
\hspace{-1cm} \; w(z) = C_1 (1-z)^{-a} {}_2F_1\left(a, c-b; a-b+1; \frac{1}{1-z}\right) + C_2 (1-z)^{-b} {}_2F_1\left(b,c-a;b-a+1; \frac{1}{1-z}\right). \nonumber \\\label{hyper2k-}
\end{eqnarray}
As $a =  d+ \frac{\tilde{\mu}}{2}+\frac{3}{4}$, $b =  d- \frac{\tilde{\mu}}{2}+\frac{3}{4}$ and $c = \frac{1}{2}$ in (\ref{eqk-}),  we can express the solution of (\ref{seg_hek-}) near infinity, that is $x = \infty$, as   
\begin{eqnarray}
\hspace{-1cm} \; \psi(x) &=& (1-|k|x^2)^{d}\left[C_1 (1-|k|x^2)^{-d-\frac{\tilde{\mu}}{2}-\frac{3}{4}}{}_2F_1\left(d+\frac{\tilde{\mu}}{2}+\frac{3}{4}, -d-\frac{1}{4}+\frac{\tilde{\mu}}{2}; 1+\tilde{\mu}; \frac{1}{1-|k| x^2}\right) \right. \nonumber \\
\hspace{-1cm} \; & & \left. + C_2 (1- |k|^2)^{-d+\frac{\tilde{\mu}}{2}-\frac{3}{4}}{}_2F_1\left(d-\frac{\tilde{\mu}}{2}+\frac{3}{4}, -d-\frac{\tilde{\mu}}{2}+\frac{3}{4}; 1-\tilde{\mu}; \frac{1}{1-|k|x^2}\right)\right]. \label{scat-eq}
\end{eqnarray}

Now we use the relation of associated Legendre polynomial $P^{m}_{\nu}(y)$ with hypergeometric function,
\begin{eqnarray}
\hspace{-1cm} \; P^{m}_{\nu}(y) &=& \frac{\Gamma\left(-\nu - \frac{1}{2}\right)\;(y^2 - 1)^{-\frac{1+\nu}{2}}}{2^{\nu+1}\sqrt{\pi}\;\Gamma{(-\nu+m)}} {}_2F_1\left(\frac{\nu+m+1}{2}, \frac{\nu-m+1}{2}; \nu+\frac{3}{2}; \frac{1}{1-y^2}\right) \nonumber \\
\hspace{-1cm} \; & & +\frac{2^{\nu}\Gamma\left(\nu + \frac{1}{2}\right)\;(y^2 - 1)^{\frac{\nu}{2}}}{\sqrt{\pi}\;\Gamma{(\nu+m+1)}} {}_2F_1\left(\frac{-\nu-m}{2}, \frac{-\nu+m}{2}; -\nu+\frac{1}{2}; \frac{1}{1-y^2}\right), \nonumber \\ 
\hspace{-1cm} & & \hspace{11cm} |1-y^2|>1  
\end{eqnarray}
and express the solution (\ref{scat-eq})  in terms of associated Legendre polynomial  in which we have chosen the arbitrary constants ${\displaystyle C_1 = \frac{\Gamma\left(-\nu - \frac{1}{2}\right)\;}{2^{\nu+1}\sqrt{\pi}\;\Gamma{(-\nu+m)}}}$, ${\displaystyle C_2 = \frac{2^{\nu}\Gamma\left(\nu + \frac{1}{2}\right)\;}{\sqrt{\pi}\;\Gamma{(\nu+m+1)}}}$  and $\nu = \tilde{\mu} - \frac{1}{2}$ and $-m = 2d+1$ to be 
\begin{eqnarray}
\psi(x) = N\;(1-|k|x^2)^{-1/2}\;P^{-2d-1}_{\tilde{\mu}-\frac{1}{2}}(\sqrt{|k|}x), \qquad |x|>\frac{1}{\sqrt{|k|}}. \label{psik-1}
\end{eqnarray}
Since the Hamiltonian is Hermitian, we consider $2d = -1+ i \rho$, where $\rho$ is a real parameter and the  solution  (\ref{psik-1}) turns out to be 
\begin{eqnarray}
\psi_{\rho}(x) = (1-|k|x^2)^{-1/2}\;P^{-i\rho}_{\tilde{\mu}-\frac{1}{2}}(\sqrt{|k|}x) \Theta(\pm\sqrt{|k|}x - 1), \label{psik-2}
\end{eqnarray}
with energy eigenvalues, 
\begin{eqnarray}
E_{\rho} = \left(\frac{\rho^2 + \mu^2 + 1}{2}+2\;\eta_2\right)\hbar^2 |k|, 
\end{eqnarray}
where $\Theta(\pm\sqrt{|k|}x - 1)$ is known as Heaviside step function, which is defined as $\Theta(y) = 1$ for $y > 0$ and $\Theta(y) = 0$ for $y < 0$. 

Here, we have considered the most general ordered form (\ref{geo1d}) that relates the hermitian ordering with the non-hermitian ordered form (\ref{hermitian}). One can obtain the most general solutions for the position dependent mass quantum systems of interest by considering the general ordered form (\ref{geo1d}). If the Schr\"{o}dinger equation corresponds to hermitian ordered form of the Hamiltonians, it results in square integrable functions on the configuration space $\mathbb{R}$ with respect to the  measure $dx$, ${\psi} \in \mathbb{L}^2(\mathbb{R}, dx)$, and the set of state functions ${\psi}$ becomes square integrable and form a Hilbert space, ${\cal H}$.  We can then obtain the eigenfunctions directly through the relation (\ref{hermitian}), as $\phi = m^{-\eta}\;\psi$, which are well defined in the space $\mathbb{L}^2(\mathbb{R}, d\mu)$. It means that the eigenfunctions $\psi$ are square integrable with respect to $dx$ and form the Hilbert space ${\cal H}$, while the state functions $\phi$ are square integrable with the measure $m^{2\eta} dx$ and form the Hilbert space ${\cal H}'$, which is isomorphic to the Hilbert space ${\cal H}$.

\subsection{Quantum solvability of the nonlinear system $H_2$}
Let us next consider the Hamiltonian, 
\begin{equation}
\hspace{3cm}H_2 = (1 + k x^2)^2 \frac{p^2}{2}+ \frac{\omega^2_0 x^2}{2 (1 + k x^2)^2}.  
\hspace{6cm}\nonumber(\ref{non-ham-nlo})
\end{equation}
Here, we consider the expanded form of non-Hermitian ordered Hamiltonian (\ref{geo1d}) instead of Hermitian ordered form (\ref{geham_heo1d}) to get normalizable and bounded solution,
\begin{equation}
\hat{H} = -\frac{\hbar^2}{2m}\left[\frac{d^2}{dx^2} + (\bar{\gamma}-\bar{\alpha}-1)\frac{m'}{m} + \left(\bar{\gamma}\frac{m''}{m} - (\bar{\alpha \gamma} + 2 \bar{\gamma})\frac{m'^2}{m^2}\right)\right] + V(x). 
\label{geham-non}
\end{equation}
 The associated time-independent one dimensional generalized Schr\"{o}dinger equation is 
\begin{eqnarray}
\hspace{-1cm} \; \Phi{''} + \frac{4 k x}{(1+kx^2)}(1+\bar{\alpha}-\bar{\gamma}) \Phi{'} + 
\left(\frac{4 \sigma_1 k}{(1 + k  x^2)}+ \frac{\frac{2 E}{\hbar^2}+ 4 \sigma_2 k}{(1 + k x^2)^2} - \frac{\omega_0^2 x^2}{\hbar^2\;(1 + k x^2)^4} \right)\Phi  = 0,\nonumber \\ 
\label{seg_hen}
\end{eqnarray}
where 
\begin{eqnarray}
\sigma_1 &=& -4 \bar{\alpha\gamma} - 3 \bar{\gamma}, 
\label{sigma1}\\
\sigma_2 &=& 4 \bar{\alpha\gamma} + 2 \bar{\gamma}. 
\label{sigma2}
\end{eqnarray}
Here we use the form of the ansatz as, 
\begin{equation}
\Phi(x) = \exp{\left(\frac{-\mu\;k\;x^2}{2\;(1+kx^2)}\right)} (1+ k x^2)^d\;(k x^2)^l\;S(x), 
\label{asy2}
\end{equation}
which reduces Eq. (\ref{seg_hen}) into 
\begin{eqnarray}
\hspace{-1cm} \; S^{''}(x) + \left(\frac{4 l}{x}+\frac{4 k (d+\bar{\alpha}-\bar{\gamma}+1) x}{1+kx^2}-\frac{2 \mu k x}{(1+kx^2)^2}\right) S^{'}(x) + 
\left(\frac{2 l(2l-1)}{x^2}+\frac{4 {\sigma} k}{1 + k  x^2}+ \frac{4 \epsilon k}{(1 + k x^2)^2}\right.\nonumber\\
\left.+\frac{4\mu (d+\bar{\alpha}-\bar{\gamma}) k }{(1 + k  x^2)^3}\right) S(x) = 0, 
\label{seg_nonx}
\end{eqnarray}
where
\begin{eqnarray}
{\sigma} &=& \sigma_1+d\left(d+\frac{3}{2}+2l+ 2\bar{\alpha} - 2 \bar{\gamma}\right)+2l(1 +\bar{\alpha}-\bar{\gamma}),\label{sigma}\\
\epsilon &=&  \frac{E}{2\;\hbar^2 k^2} + \sigma_2 -\frac{\mu}{4} (1+4l+4d+\bar{\alpha}-\bar{\gamma}) - d(d+1+2\bar{\alpha}-2\bar{\gamma}) \label{epsilon}
\end{eqnarray}
with ${\displaystyle \mu = \frac{\omega_0}{\hbar\;k}}$. The transformation ${\displaystyle \sqrt{k} x = \frac{\sqrt{z}}{\sqrt{1-z}}}$ reduces equation (\ref{seg_nonx}) into
\begin{eqnarray}
\hspace{-1cm} \; z(1 - z)^2\;S^{''}(z) + \left[2l+\frac{1}{2}+\left(2d+2\bar{\alpha} - 2 \bar{\gamma}-\mu-2l-\frac{1}{2}\right)z
+ \left(-2d- 2\bar{\alpha} + 2 \bar{\gamma}+ 2\mu\right) z^2\right.\nonumber \\
\hspace{-1cm} \left. -\mu z^3\right]S^{'}(z)+ \left[\epsilon+\sigma+\mu (d+\bar{\alpha} - \bar{\gamma}) - (\epsilon+2 \mu (d+\bar{\alpha} - \bar{\gamma})) z + \mu (d+\bar{\alpha}-\bar{\gamma}) z^2\right]S(z) = 0, \; |z| < 1.
\nonumber \\ 
\label{psiz-non}
\end{eqnarray}

This is of the form of confluent Heun equation which can be solved using the Bethe ansatz method \cite{zhang, quesne2018}. 

\subsubsection{\bf Bethe-Ansatz method} 
To start with, we first give a brief description of the method which is a quasi-exact treatment and then use
the results to solve  equation (\ref{psiz-non}), \cite{zhang, quesne2018}. Consider the differential equation of the form, 
\begin{eqnarray}
\sum^{4}_{j = 0} a_j z^j S''(z) + \sum^{3}_{j = 0} b_j z^j S'(z) + \sum^{2}_{j = 0} c_j z^j S(z) = 0, 
\label{bethe}
\end{eqnarray}
where $a_0, a_1,\;a_2, a_3,\;a_4, b_0, b_1,\;b_2,\; b_3$, $c_0, c_1$ and $c_2$ are parameters. 

Eq. (\ref{bethe}) has a $n$-degree  polynomial solution, 
\begin{equation}
S(z) = \Pi^{n}_{i = 1} (z - z_i), \qquad S(z) = 1 \quad \text{for}\quad n = 0, \label{st}
\end{equation}
with the distinct roots $z_1, z_2, . . . , z_n$, satisfying the Bethe-ansatz equations, 
\begin{equation}
\sum^n_{j\neq i} \frac{2}{z_i - z_j}= -\frac{b_3 z_i^3+b_2 z_i^2 + b_1 z_i + b_0}{a_4 z_i^4 + a_3 z_i^3+a_2 z_i^2 + a_1 z_i + a_0},\label{bethe-ansatz}
\end{equation}
provided the following restrictions on the parameters hold:  
\begin{eqnarray}
\hspace{-1cm} \quad c_2 &=& -n(n-1)a_4 - n b_3, \label{con1}\\
\hspace{-1cm} \quad c_1 &=& -n b_2 - n(n-1)a_3 - (2(n-1)a_4 + b_3)\sum^{n}_{i = 1}z_i , \label{con2}\\
\hspace{-1cm} \quad -c_0 &=&(2(n-1)a_4 + b_3)\sum^{n}_{i = 1}z_i^2 + 2 a_4 \sum^{n}_{i<k}z_i z_k + (2 (n-1) a_3+ b_2) \sum^{n}_{i = 1} z_i + n\;b_1. \label{con3}
\end{eqnarray}

On comparing Eq. (\ref{psiz-non}) with (\ref{bethe}) with $l=0$ and ${\displaystyle l = \frac{1}{2}}$, we have 
$a_0= a_4 = 0, a_1=1, a_2 = -2, a_3 = 1$ and  $b_0=2 l+ \frac{1}{2}, b_1 = 2 d +2\bar{\alpha} - 2\bar{\gamma} -\mu - 2 l -\frac{1}{2} ,\;b_2 = 2\mu -2 d -2 \bar{\alpha} + 2 \bar{\gamma},\; b_3 =-\mu$ and 
$c_0 = \epsilon+\sigma+\mu (d + \bar{\alpha} - \bar{\gamma}), \; c_1 = -\epsilon-2\mu (d+\bar{\alpha} - \bar{\gamma}), c_2 = \mu (d+\bar{\alpha} - \bar{\gamma})$. 
The relations (\ref{con1}), (\ref{con2}) and (\ref{con3}) ensure that 
\begin{eqnarray}
d &=& n +  \bar{\gamma} - \bar{\alpha}, \label{dn}\\
\epsilon&=& - \mu \sum^{n}_{i = 1} z_i - n (n+1), \label{txin}\\
\sigma&=& \mu \sum^{n}_{i} z^2_i+(2 - \mu )\sum^{n}_{i = 1} z_i-n\left(n-2 l- \frac{3}{2}\right)
\end{eqnarray}
with 
\begin{equation}
\hspace{-1cm} \qquad \sum^n_{j\neq i} \frac{2}{z_i - z_j}= \frac{\mu z_i^2 + \left(-\mu + 2 n\right) z_i +2 l + \frac{1}{2}}{z_i(z_i-1)}.\label{bethe-ansatz-non}
\end{equation}

Hence we can write down the eigenfunctions, $\Phi_n(x)$, by using (\ref{st}) in $S(z)$ and then the resultant expression in  (\ref{asy2}) with ${\displaystyle z = \frac{k x^2}{1 + k x^2}}$ as 
\begin{eqnarray}
\hspace{-1cm} \quad \Phi^{(l)}_n(x) =N^{(l)}_n \exp{\left(-\frac{\omega_0\;x^2}{2\hbar(1+k x^2)}\right)} (1 + k x^2)^{n+\bar{\gamma}-\bar{\alpha}} (k x^2)^{l}\;\; \Pi^{n}_{i = 0} \left(\frac{k x^2}{1 + k x^2} - z_i\right)
\label{non-psinx}
\end{eqnarray}
and the energy eigenvalues are 
\begin{eqnarray}
\hspace{-1cm} \quad E^{(l)}_n =\left(2n +2l+ \frac{1}{2}\right)\hbar \omega_0 +\left(\sigma_1+ \bar{\gamma}-(\bar{\gamma}-\bar{\alpha})(\bar{\gamma}-\bar{\alpha}-1)-\mu\sum^{n}_{i = 1} z_i  \right)\;2\hbar^2 k \label{non-enf}
\end{eqnarray}
with 
\begin{equation}
\sigma_1 = \mu \sum^{n}_{i} z^2_i + (2- \mu)\sum^{n}_{i = 1} z_i-2n^2 -2l+(\bar{\gamma}-\bar{\alpha})\left(\bar{\gamma}-\bar{\alpha}-\frac{3}{2}\right). 
\label{sigman1}
\end{equation} 
Here it is also noted that the term $\sigma_1$ containing ordering parameters is related to the quantum number $n$ given in (\ref{sigman1}) and hence the ordering parameters, $\alpha_i +\beta_i +\gamma_i = -1,\;i =1, 2, 3, ... N$, are considered to be arbitrary. 

{\bf (a) Ground state: $n =0$ and $l =0$}\\
When $l =0$, the ground state solution, $\Phi^{(0)}_0$, can be explicitly expressed as 
\begin{eqnarray}
\Phi^{(0)}_0(x) = N^{(0)}_{0} \exp{\left(-\frac{\omega_0\;x^2}{2\hbar(1+k x^2)}\right)}(1+k x^2)^{\bar{\gamma}-\bar{\alpha}} \label{phix00}
\end{eqnarray}
and the energy eigenvalues are
\begin{eqnarray} 
E^{(0)}_0 = \frac{\hbar\;\omega_0}{2} + \left(\bar{\alpha}+\bar{\gamma}\right)\hbar^2 k \label{ground0}
\end{eqnarray}
with a restriction on ordering parameters through (\ref{sigman1}) as 
\begin{equation}
\hspace{-1cm} \quad \bar{\alpha \gamma} = -(\bar{\gamma}-\bar{\alpha})^2-\frac{3}{2}(\bar{\gamma}+\bar{\alpha}). 
\label{sigmac1}
\end{equation}
{\bf (i)\; Boundness:}

For $k>0$, $\psi^{(0)}_0(x) = 0$ at $x = \pm \infty$ if $\bar{\gamma} - \bar{\alpha} < 0$, whereas for $k < 0$, $\psi^{(0)}_0(x) = 0$ at $x = \pm \frac{1}{\sqrt{|k|}}$ for all values of $\bar{\gamma} - \bar{\alpha}$. 

{\bf (ii) Normalizability:}

We have already seen that non-Hermitian ordered Hamiltonian $\hat{H}_{non}$ is related to the Hermitian ordered Hamiltonian $\hat{H}_{her}$ with respect to $m^{\eta}$ as given in (\ref{hermitian}). And the eigenfunction of $\hat{H}_{her}$, say $\psi$, is related to that of the non-Hermitian Hamiltonian $\hat{H}_{non}$, say $\Phi$, through the relation $\Phi =  m^{-\eta}\psi$. Hence we can call non-hermitian ordered Hamiltonian $\hat{H}_{non}$ as quasi-Hermitian. In general, the normalizability condition for the quasi-Hermitian Hamiltonian $\hat{H}_{non}$ is generalized to be $\langle m^{\eta}\Phi |m^{\eta}\Phi\rangle = 1$. But in our study we observe that the presence of $(1+k x^2)^{\bar{\gamma}-\bar{\alpha}} (= m^{2\eta})$ is responsible for the solution (\ref{non-psinx}) to be bounded as well as normalizable. 
Hence we discuss about the normalizability of the ground state solution for both the cases. For $k < 0$, the normalization condition becomes 
\begin{eqnarray}
1 &=& \int \Phi^{{(0)}^{*}}_0 \Phi^{(0)}_0 dx \nonumber \\
 & = & \int^{\frac{1}{\sqrt{|k|}}}_{-\frac{1}{\sqrt{|k|}}}\exp{\left(-\frac{\omega_0\;x^2}{\hbar(1-|k| x^2)}\right)}(1-|k| x^2)^{2\bar{\gamma}-2\bar{\alpha}}dx. \label{intg0} 
\end{eqnarray} 
On using the transformation $z = \frac{\sqrt{|k|}x}{\sqrt{1 - |k| x^2}}$, Eq. (\ref{intg0}) becomes 
\begin{eqnarray}
1 & = & 2 \frac{N^{{(0)}^2}_{0}}{\sqrt{|k|}}\int^{\infty}_{0}\exp{\left(-\frac{\omega_0\;z^2}{\hbar|k|}\right)}
(1+z^2)^{2\bar{\alpha}-2\bar{\gamma}-\frac{3}{2}}dz. \label{intgz0} 
\end{eqnarray} 
By considering $\bar{\alpha} - \bar{\gamma} = \frac{3}{4}$, the integral (\ref{intgz0}) leads to   
\begin{eqnarray}
N^{(0)}_0 &=&\left(\frac{\omega_0}{\hbar\;\pi}\right)^{1/4}.  
\end{eqnarray}
Similarly for $k>0$ with $\bar{\alpha} - \bar{\gamma} = \frac{3}{4}$, by using the transformation 
${\displaystyle z = \frac{\sqrt{k}x}{\sqrt{1 + k x^2}}}$, we can evaluate the normalization constant, 
\begin{eqnarray}
N^{(0)}_0 &=&\left(\frac{\omega_0}{\hbar\;\pi}\right)^{1/4}\frac{1}{\sqrt{erf{\left(\sqrt{\frac{\omega_0}{\hbar k}}\right)}}}, \label{noo} 
\end{eqnarray}
where $erf(a)$ is the error function. Here, we may choose $\bar{\alpha} - \bar{\gamma} = s+\frac{3}{4}, \; $ where  $s$ is an integer. For the ground state, we have chosen $s = 0$. Essentially, this choice leads to one more constraint on the ordering parameters, that is $\bar{\alpha} - \bar{\gamma} = s+\frac{3}{4}, \; $ $s = 0, 1, 2, 3,...$, besides $\bar{\alpha}+\bar{\beta}+\bar{\gamma} = -1$. Hence the solutions (\ref{phix00}) and (\ref{ground0}) now consist of only one arbitrary ordering parameter. 

We list out other possible bound states in the Table given in Appendix A. 
While normalizing these states, we fixed the value of $\bar{\alpha} - \bar{\gamma}$. In general, the value is considered to be 
\begin{equation}
\bar{\alpha} - \bar{\gamma} = n + l + \frac{3}{4}. \label{abeta34}
\end{equation}
When $n>1$, we get the polynomial of $z_i$ of order $n+1$ which cannot be solved analytically. Hence, we explicitly evaluated the first two states of the system and so it is quasi exactly solvable.
 
\section{\label{3d} Three dimensional generalization of the nonlinear oscillators}
We extend our study to the three dimensional generalization of the two nonlinear oscillators. To start with, we consider the Lagrangian corresponding to the Higgs oscillator (see (\ref{schwingerq})), 
\begin{equation}
{L} = \frac{1}{2}\left[\frac{\dot{\bf q}^2}{(1 + k {\bf q}^2)}-\frac{ k\;({\bf q}. \dot{\bf q})^2}{(1 + k  {\bf q}^2)^2}- \omega^2_0 {\bf q}^2\right],  
\label{schwingerq-higg}
\end{equation}
where $\omega_0$ is the potential parameter, which results in the equation of motion, 
\begin{eqnarray}
\hspace{-1cm} \quad \frac{\ddot{q}_i}{(1 + k {\bf q}^2)^2} - 2 k \frac{({\bf q}.\dot{\bf q})}{(1 + k {\bf q}^2)^3}\dot{q}_i + \left[\frac{2 k^2 ({\bf q}.\dot{\bf q})^2}{(1 + k {\bf q}^2)^3} + \omega^2_0\right] q_i = 0, \qquad i = 1, 2, 3. 
\label{eom-h3d}
\end{eqnarray}
We express the equation of motion (\ref{eom-h3d}) in polar coordinates $(r, \theta, \phi)$, as  
$q_1 = r \sin{\theta}\cos{\phi}, \; q_2  = r \sin{\theta} \sin{\phi}, \; q_3 = r \cos{\theta}$. 
The radial part (that is for $r$) and angular parts (for ($\theta,\;\phi$)) become, 
\begin{eqnarray}
\frac{\ddot{r}}{(1 + k r^2)^2} - \frac{2 k r}{(1 + k r^2)^3} \dot{r}^2 + \omega^2_0 r - \frac{r}{(1 + k r^2)^2}
\left(\dot{\theta}^2 + \sin^2(\theta) \dot{\phi}^2\right) = 0,\label{rhigg}\\
\ddot{\theta} + \frac{2 \dot{r} \dot{\theta}}{r\;(1 + k r^2)} - \sin{\theta}\cos{\theta} \dot{\phi}^2 = 0, \label{0higg}\\
\frac{r^2\;\sin^2{\theta}\;\dot{\phi}}{(1 + k r^2)} = C_1, \label{phigg}
\end{eqnarray}
where $C_1$ is constant. On substituting (\ref{phigg}) in (\ref{0higg}) and (\ref{rhigg}), we obtain 
\begin{eqnarray}
\frac{r^4 \dot{\theta}^2}{2\;(1 + k r^2)^2} + \frac{C^2_1}{2 \sin^2(\theta)} = \frac{C^2_2}{2}, \label{0ehigg}\\
\frac{\ddot{r}}{(1 + k r^2)^2} - \frac{2 k r \dot{r}^2}{(1 + k r^2)^3} + \omega^2_0 r - \frac{C^2_2}{r^3} = 0. \label{rehigg}
\end{eqnarray}
Equation (\ref{rehigg}) can be expressed as 
\begin{eqnarray}
\frac{\dot{r}^2}{(1 + k r^2)^2}+ \frac{C^2_2\;(1+k r^2)}{r^2} + \omega^2_0 r^2 = C_3, \label{rehigg1}
\end{eqnarray}
where $C_3$ is a third integration constant.

\begin{figure}[!ht]
\vspace{0.5cm}
\begin{center}
\includegraphics[width=0.5\linewidth]{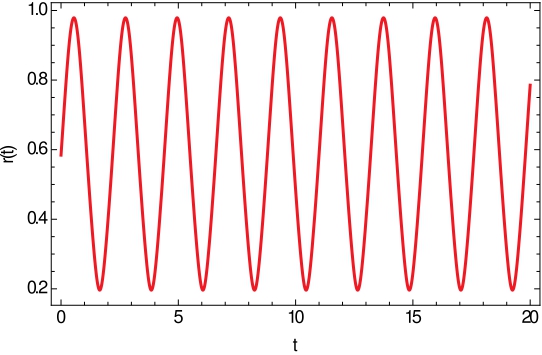}
\end{center}
\vspace{-0.3cm}
\caption{The plot of $r(t)$ Eq. (\ref{sol-r})  for $C_3 = 1,\; C_2 = 0.2,\; k = 1, \; \kappa = 0$ and $\omega_0 = 1$.} \label{first-3d}
\vspace{-0.3cm}
\end{figure}

We first consider the radial part (\ref{rehigg1}). By using the transformation, 
\begin{equation}
z = \frac{r^2}{1 + k r^2},\label{zr}
\end{equation}
Eq. (\ref{rehigg}) becomes
\begin{equation}
\dot{z}^2 = -a z^2 + b z - c, 
\label{zeq}
\end{equation}
where ${\displaystyle a = 4\left(k^2\;C^2_2 + k C_3 + \omega_0^2\right)}$, ${\displaystyle\; b = 4\;(C_3 + 2 k C^2_2)}$ and $c = 4 C^2_2$. On further using the 
transformation, $z = z'- \tau$, equation (\ref{zeq}) becomes 
\begin{equation}
\dot{z'} = \sqrt{a} \sqrt{{\Lambda}^2 - z^2}, \label{ezp}
\end{equation}
where ${\displaystyle \tau = - \frac{C_3 + 2 k C_2^2}{2 \left(k^2\;C^2_2 + k C_3 + \omega_0^2\right)}}$ and ${\displaystyle \Lambda = \frac{\sqrt{C^2_3 - 4 \omega^2_0 C^2_2}}{2 \left(k^2\;C^2_2 + k C_3 + \omega_0^2\right)}}$. 
On solving Eq. (\ref{ezp}), we can get
\begin{equation}
z'(t) = \Lambda \sin{(\Omega t + \kappa)}
\end{equation}
so that using (\ref{zr}) we obtain 
\begin{equation}
r(t) =  A \left[\frac{\eta + \sin(\Omega t + \kappa) }{1 - k \eta A^2 - k A^2 \sin(\Omega t + \kappa)}\right]^{1/2}, 
\label{sol-r}
\end{equation}
where $\kappa$ is a constant and 
\begin{eqnarray}
\Omega &=& 2 \sqrt{\left(k^2\;C^2_2 + k C_3 + \omega_0^2\right)}, \nonumber \\
A^2 &=& \Lambda = \frac{\sqrt{C^2_3 - 4 \omega^2_0 C^2_2}}{2 \left(k^2\;C^2_2 + k C_3 + \omega_0^2\right)},\nonumber \\
\eta &=& -\frac{\tau}{\Lambda} = \frac{C_3 + 2 k C_2^2}{\sqrt{C^2_3 -  4 \omega^2_0 C^2_2}}.
\end{eqnarray}
When $k>0$, the solution (\ref{sol-r}) is periodic for $C^2_3 - 4 \omega^2_0 C^2_2>0$ and $C_3 > 0$ which is shown in the figure \ref{first-3d}. When $k<0$, the periodic solutions exist for $C^2_3 - 4 \omega^2_0 C^2_2>0$ and $\Omega>0$.

Now we consider the Lagrangian corresponding to the nonlinear oscillator (via (\ref{schwingerq})), 
\begin{equation}
{L} = \frac{1}{2}\left[\frac{\dot{\bf q}^2}{(1 + k {\bf q}^2)} - \frac{k ({\bf q}. \dot{\bf q})^2}{(1 + k  {\bf q}^2)^2}- \frac{\omega^2_0 {\bf q}^2}{(1 + k \; q^2)^2}\right],  
\label{schwingerq-nlor}
\end{equation}
where $\omega_0$ is the potential parameter, which results in the equation of motion, 
\begin{eqnarray}
\hspace{-1cm} \; \frac{\ddot{q}_i}{(1 + k {\bf q}^2)^2} - 2 k \frac{({\bf q}.\dot{\bf q})}{(1 + k {\bf q}^2)^2}\dot{q}_i + \left[\frac{2 k^2 ({\bf q}.\dot{\bf q})^2}{(1 + k {\bf q}^2)^3} -\frac{ k\;({\bf q}. \dot{\bf q})}{(1 + k  {\bf q}^2)^2}+ \frac{\omega^2_0 (1 - k  {\bf q}^2)}{(1 + k  {\bf q}^2)^3}\right] q_i = 0. 
\label{eom-nlo}
\end{eqnarray}
We again express the equation of motion (\ref{eom-nlo}) in polar coordinates $(r, \theta, \phi)$ as   before. 
The radial part and angular parts become, 
\begin{eqnarray}
\ddot{r} - \frac{2 k r}{(1 + k r^2)} \dot{r}^2 + \omega^2_0 r \left( \frac{1 - k r^2}{1 + k r^2}\right)-r
\left(\dot{\theta}^2 + \sin^2{\theta} \dot{\phi}^2\right) = 0,\label{rnlo2}\\
\ddot{\theta} + \frac{2 \dot{r} \dot{\theta}}{r\;(1 + k r^2)} - \sin{\theta}\cos{\theta} \dot{\phi}^2 = 0, \label{0nlo2}\\
\frac{r^2\;\sin^2{\theta}\dot{\phi}}{(1 + k r^2)} = C_1, \label{pnlo2}
\end{eqnarray}
where $C_1$ is constant. On substituting (\ref{pnlo2}) in (\ref{0nlo2}) and (\ref{rnlo2}), we can get 
\begin{eqnarray}
\frac{r^4 \dot{\theta}^2}{2\;(1 + k r^2)^2} + \frac{C^2_1}{2 \sin^2{\theta}} = \frac{C^2_2}{2}, \label{0enlo}\\
\frac{\ddot{r}}{(1 + k r^2)^2} - \frac{2 k r \dot{r}^2}{(1 + k r^2)^3} + \omega^2_0 r \left( \frac{1 - k r^2}{1 + k r^2}\right)- \frac{C^2_2}{r^3} = 0. \label{renlo}
\end{eqnarray}
On simple transformation Eq. (\ref{renlo}) becomes that of a quadratic polynomial. It will in principle lead to elliptic functions, though complicated to express explicitly. So we do not write the form here. Therefore, we will study the quantum solvability of the nonlinear system (\ref{schwingerq-nlor}) in Sec. V C.  
 
\subsection{{\label{semihigg3d} Semiclassical method}}
The conjugate momentum for the Lagrangian (\ref{schwingerq-higg}) is 
\begin{equation}
{p}_i = \frac{\partial L}{\partial \dot{q}_i} = \frac{\dot{q}_i}{(1 + k {\bf q}^2)} - \frac{k ({\bf q}. {\bf \dot{q}}) q_i}{(1 + k {\bf q}^2)^2} \label{momentum-higg}
\end{equation}
and 
\begin{eqnarray}
{\bf p}.{\bf p} = \frac{{\bf \dot{q}}^2}{(1 + k {\bf q}^2)^2} - \frac{k\;(2-k {\bf q}^2)\; ({\bf q}.{\bf \dot{q}})^2}{(1 + k {\bf q}^2)^4}, \qquad
{\bf p}.{\bf q} = \frac{{\bf q}.{\bf \dot{q}}}{(1 + k {\bf q}^2)^2}
\end{eqnarray}
yield the Hamiltonian corresponding to the Lagrangian (\ref{schwingerq-higg}) as 
\begin{equation}
H = \frac{1}{2}\left[(1 + k {\bf q}^2)({\bf p}^2 + k ({\bf p}. {\bf q})^2) + \frac{\omega_0^2\; {\bf q}^2}{2} \right]. 
\label{ham-higg}
\end{equation}

In polar coordinates, the Hamiltonian can be expressed as 
\begin{equation}
H = \frac{1}{2}\left[\frac{{\dot{r}^2}}{(1 + k r^2)^2} +\frac{r^2\;({\dot{\theta}^2 + \sin^2{\theta}\;{\dot{\phi}^2}})}{(1 + k r^2)} + \omega^2_0 r^2 \right]. 
\label{ham}
\end{equation}

On substituting (\ref{pnlo2}) and (\ref{0enlo}), we can reduce the Hamiltonian (\ref{ham}) as 
\begin{equation}
H = \frac{1}{2}\left[\frac{{\dot{r}^2}}{(1 + k r^2)^2} +\frac{C^2_2 (1 + k r^2)}{r^2} + \omega^2_0 r^2 \right]. 
\label{ham-c2}
\end{equation}

Now using the solution (\ref{sol-r}), the energy for the classical system (\ref{ham-higg}) can be obtained as 
\begin{equation}
E = \frac{1}{2}\left[\frac{\Omega^2}{4 k} - \frac{\omega^2_0}{k} \right]. 
\label{ham-c2}
\end{equation}

To obtain the semiclassical energy eigenvalues, we apply the Bohr-Sommerfeld quantization rule (\ref{bohr}) to the radial part, 
\begin{equation}
\oint{p_r dr} = \left(n_r + \frac{1}{2}\right) h = I, \quad n_r =0, 1, 2, .... \label{nr_bohr}
\end{equation}
The momentum $p_r$ can be expressed as 
\begin{equation}
p_r = \frac{\dot{r}}{(1 + k r^2)^2},  
\label{mom_r}
\end{equation}
along with the quantization of angular parts 
\begin{equation}
\oint{p_{\theta}} d{\theta} = l h, \quad l = 0, 1, 2, ...n . \label{ntheta_bohr}
\end{equation}

On substituting the solution (\ref{sol-r}) in (\ref{mom_r}) and using (\ref{rehigg1}) and (\ref{ham-c2}) in the integral (\ref{nr_bohr}), we can get
\begin{eqnarray}
I &=& \frac{1}{\Omega}\int^{2\;\pi}_{0} \left(C_3 + \frac{\omega^2_0}{k} - \frac{C^2_2}{A^2\;(\eta + \sin(u))}
-\frac{\omega^2_0}{k\;(1 - k\eta A^2- k A^2 \sin(u))}\right) du, \nonumber\\
  &=& \frac{2\;\pi}{\Omega}\left(C_3 + \frac{\omega^2_0}{k}- \frac{C^2_2}{A^2\;\sqrt{\eta^2 - 1}}-\frac{\omega^2_0}{k\;\sqrt{(1 -  k A^2 \eta)^2 - k^2 A^4)}}\right),  \label{int-3d}  
\end{eqnarray}
where $u = \Omega t + \kappa.$

On evaluating (\ref{int-3d}), we can get 
\begin{equation}
\frac{\Omega}{2 k} -C_2 - \frac{\omega_0}{k} = (2 n_r + 1)\hbar,  \label{eneq}
\end{equation}
which sets the energy value (\ref{ham-c2}) as 
\begin{equation}
E_n = \left(2 n_r + l + \frac{3}{2}\right) \hbar \omega_0 + \left(2 n_r + l + \frac{3}{2} \right)^2 
\frac{\hbar^2 k}{2}. \label{energy-semi}
\end{equation}
The semiclassical energy is also of the form of a quadratic in $n_r$. It confirms that the quantum counterpart of the system (\ref{ham-higg}) can also be exactly solved. 

We do not here apply the semiclassical method to  the system  (\ref{schwingerq-nlor}) as one requires the explicit form of the classical solution of the system,  as it is quite cumbersome as pointed out earlier. However, we will carry out the Bethe ansatz method to this system in the following.

\subsection{{\label{qhigg3d} Quantum solvability of the three dimensional Higgs oscillator}}
The components of conjugate momentum in polar coordinates $(p_r, p_{\theta}, p_{\phi})$ can be obtained as  
\begin{equation}
p_r = \frac{\dot{r}}{(1 + k r^2)^2}, \quad p_{\theta} = \frac{r^2}{(1 + k r^2)}\dot{\theta}, \quad {\text and}\quad  p_{\theta} = \frac{r^2\;\sin^2{\theta}}{(1 + k r^2)}\dot{\phi}. 
\label{mom}
\end{equation}
On substituting (\ref{mom}) in (\ref{ham}), we can get 
\begin{equation}
H = \frac{1}{2}\left[(1 + k r^2)^2 p_r^2 +\frac{(1 + k r^2)}{r^2}\left(p^2_{\theta} + \frac{p^2_{\phi}}{\sin^2{\theta}}\right) + \omega^2_0 r^2 \right]. 
\label{ham3dr}
\end{equation}
The Hamiltonian (\ref{ham3dr}) consists of position ($r$) term along with momentum term $p_r$ and  obviously corresponds to a position-dependent mass (PDM) quantum system. While quantizing the Hamiltonian, it is required that the ordering between momentum and mass operators in the kinetic energy term is taken care of appropriately. In this three dimensional case also, we consider the most general ordered form of the Hamiltonian operator, 
\begin{eqnarray}
\hat{H} = \frac{1}{2}\left[\sum^N_{i = 1} w_i m^{\alpha_i} \hat{p}_r m^{\beta_i} \hat{p}_r m^{\gamma_i}+\frac{(1 + k r^2)}{r^2}\left(p^2_{\theta} + \frac{p^2_{\phi}}{\sin^2{\theta}}\right) + \omega^2_0 r^2 \right],
\label{geo}
\end{eqnarray}
where  $N$ is an arbitrary positive integer, and the ordering parameters should satisfy the constraint $\alpha_i +\beta_i +\gamma_i = -1,\;i =1, 2, 3, ... N$ and $w_i$'s  are real weights which are summed to be $1$. 
The  Hermitian Hamiltonian  (\ref{hermitian})  for a potential $V(r) = \frac{\omega^2_0\;r^2}{2}$ with mass ${\displaystyle m(r) = \frac{1}{(1 + k r^2)^2}}$ can be expressed as, 
\begin{eqnarray}
\hspace{-1cm} \; \hat{H}_{her} = \frac{1}{2}\left(\frac{1}{m}\hat{p}^2_r - i \hbar \frac{d}{d\;r}\left(\frac{1}{m}\right)\hat{p}_r\right) + \frac{\hbar^2}{2}\left[\left(\frac{\bar{\alpha}+\bar{\gamma}}{2}\right) \frac{d^2}{d r^2}\left(\frac{1}{m}\right) + \left(\overline{\alpha\gamma}+ \frac{1}{4} (\bar{\gamma} - \bar{\alpha})^2\right)  \left(\frac{d}{dr}\left(\frac{1}{m}\right)\right)^2 m\right.\nonumber\\
\left. + \frac{1}{\sqrt{m}\;r^2}\left(\hat{p}^2_{\theta} + \frac{\hat{p}^2_{\phi}}{\sin^2{\theta}}\right)\right] + V(r),  
\label{geham_higg}
\end{eqnarray}
where ${\displaystyle \hat{p}_r = - i \hbar \left(\frac{\partial}{\partial r} + \frac{1}{r}\right)},\quad {\displaystyle \hat{p}_{\theta} = - i \hbar \left(\frac{1}{\sin{\theta}} \frac{\partial}{\partial \theta}\right)}$ and ${\displaystyle \hat{p}_{\phi} = -i\hbar \frac{\partial}{\partial \phi}}$. 
The generalized Schr\"{o}dinger equation can be written as
\begin{eqnarray} 
\hspace{-1cm} \Psi''+ \left(\frac{2}{r}-\frac{m'}{m}\right)\Psi'+ \left[\left(\frac{\bar{\alpha}+\bar{\gamma}}{2}\right) \left(\frac{m''}{m}\right) - \left(\overline{\alpha\gamma}+\bar{\alpha}+\bar{\gamma}+ \frac{1}{4} (\bar{\gamma} - \bar{\alpha})^2\right)  \left(\frac{m'^2}{m^2}\right) \right.\nonumber \\
\hspace{-2cm}\left.-\frac{\hat{L}^2}{r^2\;\sqrt{m}}+ \frac{2\;m}{\hbar^2}\left(E - V(r)\right)\right]\Psi = 0,  
\label{geham_heo3d}
\end{eqnarray}
where the square of the angular momentum operator  $\hat{L}^2$ is
\begin{equation}
\hat{L}^2 =-\hbar^2\left(\frac{1}{\sin{\theta}}\frac{\partial}{\partial \theta}\left(\frac{1}{\sin{\theta}}\frac{\partial}{\partial \theta}\right)+\frac{1}{\sin^2{\theta}}\frac{\partial^2}{\partial \phi^2}\right),
\end{equation}
where $' = \frac{\partial}{\partial r}$. 
Considering the wave function to be  of the form,
${\displaystyle \Psi(r, \theta, \phi) = \frac{\chi(r)}{r} Y(\theta, \phi)}$, with a spherically symmetric factor 
${\displaystyle \frac{\chi(r)}{r}}$ with $m(r) = \frac{1}{(1 + k r^2)^2}$ and a generalized angular function $Y(\theta, \phi)$, Eq. (\ref{geham_heo3d}) can be separated as follows, 
\begin{eqnarray}
\hspace{-1cm} \;\;\mbox{Angular part:}&\;\;\;\;\;\;\;L^2 Y(\theta, \phi) = \hbar^2\;l(l+1)\; Y(\theta, \phi),& \label{a3dim}\\
\hspace{-1cm} \;\;\mbox{Radial part:}&\;\;{\displaystyle \chi''+\frac{4\;k r}{(1+k\;r^2)}\chi'+ \left[\frac{A}{(1 + k r^2)} + \frac{B}{(1+kr^2)^2} - \frac{l(l+1)}{r^2}\right]\chi = 0,} &
\label{r3dim}
\end{eqnarray}
where $Y_{l, m}(\theta, \phi) = \sqrt{\frac{(l - m)!}{(l + m)!}}\; e^{i m \phi}\; P^m_l(\cos{\theta})$ are spherical harmonics. Here $P^m_l$ is the associated Legendre polynomial, $l$ is the angular momentum quantum 
number and $m$ is the magnetic quantum number. For every value of $l$, $m = -l, -(l-1), ... 0, ... l-1, l$, that is $m$ takes $2 l + 1$ values. Here 
\begin{eqnarray}
A &=& 2\eta_1 - k\;\mu^2 + k l(l+1),\\
B &=& 4\eta_2 + k\;\mu^2 + \frac{2 E}{\hbar^2},\label{energy-3d1}\\
\mu &=& \frac{\omega_0}{\hbar\;k}.\label{muh}
\end{eqnarray}

As we did in the one dimensional case, we solve the radial part (\ref{r3dim}) for both $k > 0$ and $k<0$.
 
\subsection*{(a)\; Positive values of $k$}
When $k > 0$, $r$ spreads over from $0$ to $\infty$. Let us consider the transformation
\begin{eqnarray}
\sqrt{k}\;r =\frac{\sqrt{z}}{\sqrt{1-z}}. 
\label{rtrans}
\end{eqnarray}
Then Eq. (\ref{r3dim}) gets reduced to  
\begin{equation}
z(1 -z)\chi''(z) + \frac{1}{2} \chi'(z) + \left[\frac{B}{4\;k} + \frac{A}{4k\;(1 - z)}+ \frac{l(l+1)}{4\;z}\right] \chi(z) = 0.   
\label{chi-3d}
\end{equation}
On assuming $\chi(z) = z^{\frac{l+1}{2}}(1-z)^{\frac{3}{4}+\frac{\tilde{\mu}}{2}}\phi(z)$, Eq. (\ref{chi-3d}) reduces to 
\begin{equation}
z(1 -z)\phi''(z) + \left[l+\frac{3}{2} - \left(l+\tilde{\mu}+\frac{5}{2}\right)\right]\phi'(z) + \left[\frac{B}{4\;k} - \left(\frac{l+\tilde{\mu}}{2} +\frac{5}{4}\right)\left(\frac{l+\tilde{\mu}}{2} +\frac{1}{4}\right)\right] \phi(z) = 0.  
\label{chi-3d1}
\end{equation}
This is of the form of a hypergeometric equation which admits the solution at $z=0$ as  
\begin{equation}
\phi(z) = C_1\;{}_2{F}_{1}\left(b,-b+l+\tilde{\mu}+\frac{3}{2}; l+\frac{3}{2}; z\right) + C_2\;z^{-l-\frac{1}{2}} {}_2{F}_{1}(b-l-\frac{1}{2}, -b+\tilde{\mu}+1;\frac{1}{2}-l, z), \; |z|<1.   
\label{phi3d}
\end{equation} 
It reduces to polynomial form when $b= -n_r$, 
\begin{equation}
\phi(z) = C_1\;{}_2{F}_{1}\left(-n_r,n_r+l+\tilde{\mu}+\frac{3}{2}; l+\frac{3}{2}; z\right) + C_2 z^{-l-\frac{3}{2}} {}_2{F}_{1}(-n_r-l-\frac{1}{2}, n_r+\tilde{\mu}+1;\frac{1}{2}-l, z), \; |z| < 1,   
\label{phi3d1}
\end{equation} 
where $C_1$ and $C_2$ are integration constants, which sets  the energy eigenvalues (\ref{energy-3d1}) as  
\begin{equation}
E_{n_r,l} = \left(2 n_r + l + \frac{3}{2}\right)\hbar\omega_0 \sqrt{1+\frac{1}{\mu^2}\left(\frac{9}{4}-2\eta_1\right)}+ \left[\left(2 n_r + l+\frac{3}{2}\right)^2  + 2\bar{\alpha}+2\bar{\gamma}+\frac{5}{4}\right]\frac{\hbar^2\;k}{2}. 
\end{equation}
The boundary condition, that is at $z = 1$, $\phi_n(z) = 0$  and the continuity condition, $\phi_n(z) = \text{constant} ={\cal N}$ at $z=0$ relates the constants $C_1$ and $C_2$, if we choose  
\begin{equation}
C_1 = \frac{\Gamma(-l-\frac{1}{2})}{\Gamma(n_r + 1) \Gamma(-l - n_r -\frac{1}{2})} {\cal N}, \qquad C_2 = - \frac{\sin(n_r \pi) \Gamma(l+\frac{1}{2}) \Gamma(\mu + n_r + 1)}{\pi \Gamma(n_r+l+\mu+\frac{3}{2})}\;{\cal N}. 
\label{c1c2}
\end{equation}
Using now the identity, 
\begin{eqnarray}
\hspace{-1cm} \; P^{(a, b)}_{\nu}(x) &=& \frac{\Gamma(-b)}{\Gamma(\nu + 1) \Gamma(-b - \nu)} {}_2F_{1}\left({-\nu, a + b + \nu + 1}, {b + 1}, \frac{x + 1}{2}\right) \nonumber \\
\hspace{-1cm} \;                     & &- \frac{\sin(\nu \pi) \Gamma(b) \Gamma(a + \nu + 1)}{\pi \Gamma(a + b + \nu + 1)} {}_2F_{1}\left({a + \nu + 1, -b - \nu}; {1 - b};\frac{x + 1}{2}\right)\left(\frac{x + 1}{2}\right)^{-b}, \nonumber\\
\hspace{-1cm} \;                     & &\hspace{8cm} b \in Integers, 
\end{eqnarray}
where $P^{(a, b)}_{\nu}(x)$ are Jacobi polynomials and $\nu,\;a,\; b$ are parameters, we can reduce the solution (\ref{phi3d1}) as 
\begin{equation}
\phi_n(z) = {\cal N}_{n_r,l}\;P^{(\tilde{\mu}, l+1/2)}_{n_r}(2 z -1). 
\end{equation}
Here ${\cal N}_{n_r, l}$ is the normalization constant. 
Then the solution of (\ref{chi-3d}) becomes 
\begin{equation}
\chi(z) = {\cal N}_{n_r,l}\;z^{\frac{l+1}{2}}(1-z)^{\frac{3}{4}+\frac{\tilde{\mu}}{2}}\;N_n\;P^{(\tilde{\mu}, l+1/2)}_{n_r}(2 z -1). 
\label{fchiz}
\end{equation}
Then, in terms of $r$, 
\begin{equation}
\chi_{n,l}(r) = {\cal N}_{n,l}\;\frac{(k\;r^2)^{\frac{l+1}{2}}}{(1 + k r^2)^{\frac{l}{2}+\frac{\tilde{\mu}}{2}+\frac{5}{4}}}\;P^{(\tilde{\mu},\; l+1/2)}_{n_r}\left(\frac{k r^2 - 1}{k r^2 + 1}\right). 
\label{fchiz}
\end{equation}

The eigenfunction can also be expressed in terms of associated Jacobi polynomials as
\begin{equation}
\Psi_{n_r,l,m}(r,\theta, \phi) = {\cal N}_{n_r,l}\;\frac{1}{\sqrt{k}}\;\frac{(k\;r^2)^{\frac{l}{2}}}{(1 + k r^2)^{\frac{l}{2}+\frac{\tilde{\mu}}{2}+\frac{5}{4}}}\;P^{(\tilde{\mu},\; l+1/2)}_{n_r}\left(\frac{k r^2 - 1}{k r^2 + 1}\right)\;Y_{l,m}(\theta, \phi).  
\end{equation}

The normalization constant can be evaluated as 
\begin{eqnarray}
1 &=& \oint \psi^{*}_{n,l,m}(r,\theta, \phi)\psi_{n,l,m}(r,\theta, \phi) r^2 \sin{\theta} dr d\theta d\phi,\nonumber 
\end{eqnarray}
so that 
\begin{eqnarray}
1 &= & {\cal N}^2_{n_r,l}\int^{\infty}_{0} \int^{2\pi}_{0}\int^{\pi}_{-\pi}\frac{|\chi(r)|^2}{r^2}|Y_{l,m}|^2  r^2 \sin{\theta} dr d\theta d\phi. 
\end{eqnarray}
Hence the normalization constant becomes 
\begin{eqnarray}
{\cal N}_{n_r,l} &=& \left(\frac{2 k \sqrt{k}\;n_r! (2 n_r+\tilde{\mu} +l + 3/2) \Gamma(n_r+l+\tilde{\mu}+3/2)}{\Gamma(n_r+\tilde{\mu}+ 1) \Gamma(n_r+l+3/2)}\right)^{\frac{1}{2}}. 
\end{eqnarray}
\subsection*{(b) Negative values of $k$}
When $k < 0$, the radius is confined within the region $\left(0, \frac{1}{\sqrt{|k|}}\right)$. Hence we solve the system (\ref{r3dim}) in two regions, 
\begin{eqnarray}
\mbox{Region I}&:& 0 \leq r < \frac{1}{\sqrt{|k|}}, \\
\mbox{Region II}&:&  |r| >\frac{1}{\sqrt{|k|}},
\end{eqnarray}
We use transformation
\begin{eqnarray}
z = {|k|}\;r^2, \mbox{and} \chi(z) = z^s (1-z)^d \phi(z), 
\label{rtrans2}
\end{eqnarray}
and Eq. (\ref{r3dim}) turns out to be 
\begin{equation}
z(1 -z)\phi''(z) + \left[2s+\frac{1}{2} - \left(2s+2d+\frac{5}{2}\right)\right]\phi'(z) - \left(s+d+\frac{3}{4}+\frac{\tilde{\mu}}{2}\right)\left(s+d+\frac{3}{4}-\frac{\tilde{\mu}}{2}\right)\phi(z) = 0, 
\label{chi-3dk2}
\end{equation}
where $\tilde{\mu} = \sqrt{\frac{9}{4}+\mu^2-2 \eta_1}$ and 
\begin{eqnarray}
s\left(s+\frac{1}{2}\right) - \frac{l(l+1)}{4}  &=& 0, \label{sr}\\
\frac{E}{2\hbar^2\;|k|} - \eta_2-\frac{\mu^2}{4}+d(d+1) &=& 0.\label{er}
\end{eqnarray}
Eq. (\ref{sr}) results in $s = \frac{l+1}{2}$ and $s= \frac{l}{2}$ in which we consider $s = \frac{l+1}{2}$ as we are seeking well-defined function of $(\ref{r3dim})$. 
Eq. (\ref{chi-3dk2}) is of the form of hypergeometric equation and  admits the solution at $r =0$ (via (\ref{hyper-sol})) as
\begin{equation}
\phi(z) = C_1\;{}_2{F}_{1}(a,b; c, z) + C_2 z^{1-c} {}_2{F}_{1}(a-c+1, b-c+1; 2-c, z), \; |z|<1.  
\end{equation} 
Then the solution for (\ref{chi-3d1}) can be written as 
\begin{eqnarray}
\phi(z) = C_1\;{}_2{F}_{1}\left(l+d+\frac{5}{4}+\frac{\tilde{\mu}}{2},l+d+\frac{5}{4}-\frac{\tilde{\mu}}{2};l+\frac{3}{2}; z\right) \nonumber \\+ C_2\;z^{-l-\frac{1}{2}} {}_2{F}_{1}(d+l-\frac{3}{2}-\frac{\tilde{\mu}}{2},d+l-\frac{3}{2}+\frac{\tilde{\mu}}{2};\frac{1}{2}-l, z), \; |z|<1.   
\label{phi3dk-}
\end{eqnarray} 
It reduces to polynomial form when $d= -n_r-\frac{l}{2}-\frac{5}{4}+\frac{\tilde{\mu}}{2}$, 
\begin{eqnarray}
\hspace{-1cm} \quad \qquad \phi(z) = C_1\;{}_2{F}_{1}\left(-n_r,n_r+\tilde{\mu}; l+\frac{3}{2}; z\right) + C_2 z^{-l-\frac{3}{2}} {}_2{F}_{1}(-n_r-l-\tilde{\mu}, -n_r-l;\frac{1}{2}-l, z), \nonumber \\  \hspace{10cm} |z|<1, 
\label{phi3d1k-}
\end{eqnarray} 
where $C_1$ and $C_2$ are integration constants, which sets  the energy eigenvalues (\ref{er}) as  
\begin{equation}
E_{n_r,l} = \left(2 n_r + l + \frac{3}{2}\right)\hbar\omega_0 \sqrt{1+\frac{1}{\mu^2}\left(\frac{9}{4}-2 \eta_1\right)}- \left[\left(2 n_r + l+\frac{3}{2}\right)^2  + 2\bar{\alpha}+2\bar{\gamma}+\frac{5}{4}\right]\frac{\hbar^2\;|k|}{2}. 
\end{equation}
And the solution $\chi_{n,l}(r)$ can be expressed as, 
\begin{eqnarray}
\hspace{-1cm} \; \chi_{n,l}(r) &=& {\cal N}_{n,l}\;\frac{(|k|\;r^2)^{\frac{l+1}{2}}}{(1 - |k| r^2)^{n_r+\frac{l}{2}-\frac{\tilde{\mu}}{2}+\frac{5}{4}}}\;\left[C_1\;{}_2{F}_{1}\left(-n_r,n_r+\tilde{\mu}; l+\frac{3}{2}; |k| r^2\right) \right. \nonumber \\
\hspace{-1cm} \; & & \qquad \qquad\left.+ C_2 (|k| r^2)^{-l-\frac{1}{2}} {}_2{F}_{1}(-n_r-l-\tilde{\mu}, -n_r-l;\frac{1}{2}-l, |k| r^2)\right]. 
\label{chizk-}
\end{eqnarray}
The eigenstates are bounded if $n_r+\frac{l}{2}+\frac{5}{4} > \frac{\tilde{\mu}}{2}$ and $n_r =0, 1, 2, ...N$. 

Similarly for the region $II$, we can obtain the solution $\chi_{n,l}(r)$ for  $|r|>\frac{1}{\sqrt{|k|}}$ as
\begin{eqnarray}
\hspace{-1cm} \; \chi_{n,l}(r) &=& (|k|\;r^2)^{\frac{l+1}{2}}}{\left[C_1 (1-|k|r^2)^{-\frac{l}{2}-\frac{5}{4}+\frac{\tilde{\mu}}{2}}{}_2F_1\left(d+\frac{l}{2}-\frac{\tilde{\mu}}{2}+\frac{5}{4}, -d-\frac{l}{2}+\frac{\tilde{\mu}}{2}+\frac{1}{4}; 1-\tilde{\mu}; \frac{1}{1-|k| r^2}\right) \right. \nonumber \\
\hspace{-1cm} \; & &\left. + C_2 (1-|k|r^2)^{-\frac{l}{2}-\frac{5}{4}-\frac{\tilde{\mu}}{2}}{}_2F_1\left(d+\frac{l}{2}+\frac{\tilde{\mu}}{2}+\frac{5}{4}, -d-\frac{l}{2}+\frac{\tilde{\mu}}{2}+\frac{1}{4}; 1+\tilde{\mu}; \frac{1}{1-|k| r^2}\right) \right],  \label{scatk-}
\end{eqnarray}
where we have used (\ref{hyper2k-}). The states (\ref{scatk-}) are continuum states. 
 
\subsection{\label{nlo3d}Quantum solvability of nonlinear oscillator $V_2(r)$}
The Hamiltonian corresponding to the Lagrangian (\ref{schwingerq-nlor}) can be expressed as 
\begin{equation}
H = \frac{1}{2}\left[(1 + k {\bf q}^2)({\bf p}^2 + k ({\bf p}. {\bf q})^2) + \frac{\omega_0^2\; {\bf q}^2}{2(1+k {\bf q}^2)^2} \right]. 
\label{ham-nlo1}
\end{equation}
In polar coordinates, the Hamiltonian can be expressed as 
\begin{equation}
H = \frac{1}{2}\left[\frac{{\dot{r}^2}}{(1 + k r^2)^2} +\frac{r^2\;({\dot{\theta}^2 + \sin^2{\theta}\;{\dot{\phi}^2}})}{(1 + k r^2)} + \frac{\omega^2_0 r^2}{(1 + k r^2)^2} \right]. 
\label{ham-nlo}
\end{equation}
On substituting the components of momentum (\ref{mom}) in (\ref{ham-nlo}), we can get 
\begin{equation}
H = \frac{1}{2}\left[(1 + k r^2)^2 p_r^2 +\frac{(1 + k r^2)}{r^2}\left(p^2_{\theta} + \frac{p^2_{\phi}}{\sin^2{\theta}}\right) + \frac{\omega^2_0 r^2}{(1 + k r^2)^2} \right]. 
\label{ham3dr-nlo}
\end{equation}

To get normalizable and bounded solutions for the system (\ref{ham3dr-nlo}), we consider the non-Hermitian ordered form of the Hamiltonian (\ref{geo}),  for the potential $V(r) = \frac{\omega^2_0\;r^2}{2(1 + k r^2)^2}$ with mass ${\displaystyle m(r) = \frac{1}{(1 + k r^2)^2}}$ as, 
\begin{eqnarray}
\hspace{-1cm} \; \hat{H} &= &\frac{1}{2}\left(\frac{1}{m}\hat{p}^2_r - i \hbar (\bar{\alpha} - \bar{\gamma} + 1) \frac{d}{d\;r}\left(\frac{1}{m}\right)\hat{p}_r\right) + \frac{\hbar^2}{2}\left[\bar{\gamma}\frac{d^2}{d r^2}\left(\frac{1}{m}\right) + \overline{\alpha\gamma}  \left(\frac{d}{dr}\left(\frac{1}{m}\right)\right)^2 m\right.\nonumber\\
\hspace{-1cm} \; & &\hspace{7cm} \left. + \frac{1}{\sqrt{m}\;r^2}\left(\hat{p}^2_{\theta} + \frac{\hat{p}^2_{\phi}}{\sin^2{\theta}}\right)\right] + V(r),  
\label{geham_nlo}
\end{eqnarray}
where ${\displaystyle \hat{p}_r = - i \hbar \left(\frac{\partial}{\partial r} + \frac{1}{r}\right)},\quad {\displaystyle {\hat{p}}_{\theta} = - i \hbar \left(\frac{1}{\sin{\theta}} \frac{\partial}{\partial \theta}\right)}$ and ${\displaystyle \hat{p}_{\phi} = -i\hbar \frac{\partial}{\partial \phi}}$. 
The generalized Schr\"{o}dinger equation can be written as
\begin{eqnarray} 
\hspace{-1cm} \; \Phi''+ \left(\frac{2}{r}-(1+ \bar{\alpha} - \bar{\gamma})\frac{m'}{m}\right)\Phi'+ \left[\frac{(\bar{\gamma}-\bar{\alpha}-1)}{r}\frac{m'}{m}+\bar{\gamma} \frac{m''}{m} - \left(\overline{\alpha\gamma}+2\bar{\gamma} \right)\frac{m'^2}{m^2}\right.\nonumber \\
\hspace{-1cm} \; \hspace{7cm} \left.-\frac{\sqrt{m}\;\hat{L}^2}{r^2}+ \frac{2\;m}{\hbar^2}\left(E - V(r)\right)\right]\Phi = 0. 
\label{geham_non3d}
\end{eqnarray}
Again, we consider ${\displaystyle \Phi(r, \theta, \phi) = \frac{1}{r}\;\chi(r) Y(\theta, \phi)}$, with a spherically symmetric factor ${\displaystyle \frac{\chi(r)}{r}}$ and ${\displaystyle m(r) = \frac{1}{(1 + k r^2)^2}}$ and a generalized angular function $Y(\theta, \phi)$ which separate the equation (\ref{geham_non3d}) into 
\begin{eqnarray}
\hspace{-1cm}\;\mbox{Angular part:}&\;\;\;\;\;L^2 Y(\theta, \phi) = \hbar^2\;l(l+1)\; Y(\theta, \phi),& \label{a3dim-nlo}\\
\hspace{-1cm}\;\mbox{Radial part:}&{\displaystyle \chi''+\frac{4\;(\bar{\alpha}-\bar{\gamma}+1)\;k r}{(1+k\;r^2)}\chi'+ \left[\frac{4\sigma_1 k-kl(l+1)}{(1 + k r^2)} + \frac{\frac{2 E}{\hbar^2} + 4 \sigma_2\;k}{(1+kr^2)^2} - \frac{\mu^2\;k r^2}{(1 + k r^2)^4} - \frac{l(l+1)}{r^2}\right]\chi = 0,} &\nonumber \\
\label{r3dim-nlo1}
\end{eqnarray}
where $Y_{l, m}(\theta, \phi) = \sqrt{\frac{(l - m)!}{(l + m)!}}\; e^{i m \phi}\; P^m_l(\cos{\theta})$ are spherical harmonics, and $l,m$ are quantum numbers. 

Now we consider the quasi polynomial solution, 
\begin{equation}
\chi(r) = \exp{\left(-\frac{\mu k r^2}{2(1 + k r^2)}\right)}\;(1 + k r^2)^{d}\;(k r^2)^{s} U(r), \label{solr}
\end{equation}
so that Eq. (\ref{r3dim-nlo1}) becomes 
\begin{eqnarray}
\hspace{-1cm} \;U''(r)+\left[\frac{4 s}{r}+\frac{4\;(d+\bar{\alpha}-\bar{\gamma}+1)\;k r}{(1+k\;r^2)}-\frac{2 \mu k r}{(1+k r^2)^2}\right]U'(r)+ \left[\frac{4\sigma_r k}{(1 + k r^2)} + \frac{4\;\epsilon_r\;k}{(1+kr^2)^2} \right. \nonumber \\
\hspace{7cm} \left.+  \frac{4 \mu k (d + \bar{\alpha}-\bar{\gamma})}{(1 + k r^2)^3}\right]U(r) = 0, 
\label{r3dim-nlo}
\end{eqnarray}
where 
\begin{eqnarray}
s & =& \frac{l+1}{2}\quad \mbox{and}\qquad s = -\frac{l}{2}, \label{svalue}\\
\sigma_r & = & \sigma_1 + \frac{l(l+1)}{4} + d \left(d + \frac{3}{2} + 2 s + 2 \bar{\alpha} - 2 \bar{\gamma}\right) + 2 s (1 + \bar{\alpha} - \bar{\gamma}), \label{sigr}\\
\epsilon_r & = & \frac{E}{2\hbar^2 k} + \sigma_2 -\mu \left(d+s+\bar{\alpha}-\bar{\gamma}+\frac{1}{4}\right)-d(d+1+2\bar{\alpha} - 2\bar{\gamma})\label{epsr}. 
\end{eqnarray}

On applying again the transformation  (\ref{rtrans}), we can express Eq. (\ref{r3dim-nlo}) as 
\begin{eqnarray}
\hspace{-2cm}z(1 -z)^2\;U''(z) + \left[2s+ \frac{1}{2}+\left(2d+2\bar{\alpha} - 2 \bar{\gamma}-2s-\mu -\frac{1}{2}\right)z+(2\mu - 2d - 2\bar{\alpha}- 2 \bar{\gamma})\;z^2 \right. \nonumber \\ 
 \hspace{-2cm}\left. - \mu z^3\right] U'(z) +\left[\epsilon_r +\sigma_r+\mu (d +\bar{\alpha} - \bar{\gamma})-z(\epsilon_r + 2 \mu (d+\bar{\alpha}-\bar{\gamma}))+z^2 \mu (d+\bar{\alpha} - \bar{\gamma})\right] U(z) = 0, \nonumber \\
\hspace{-1cm}|z| < 1. 
\label{chi-3d-nlo}
\end{eqnarray}
This is again of the form of confluent Heun equation which can be solved using Bethe-ansatz method. 
Using the results of Sec. \ref{pdm} to solve equation (\ref{chi-3d-nlo}) as 
\begin{eqnarray}
U(z) = \Pi^{n}_{i=0}(z - z_i) 
\label{bethe1}
\end{eqnarray}
provided the following restrictions on the parameters hold: 
\begin{eqnarray}
d &=& n + \bar{\gamma} - \bar{\alpha}, \label{con1-nlo}\\
-\epsilon_r &=& n(n+1) + \mu\sum^{n}_{i = 1} z_i,  \label{con2-nlo}\\
-\epsilon_r - \sigma_r & = & -\mu \sum^{n}_{i = 1} z^2_i + 2(\mu - 1) \sum^{n}_{i =1} z_i + 2n^2-2ns-\frac{n}{2}. \label{con3-nlo}
\end{eqnarray}
The energy eigenvalues are 
\begin{eqnarray}
\hspace{-1cm} \; E_{n,l} &=&\left(2 n + 2s +\frac{1}{2}\right) \hbar \omega_0 +\left(-\mu \sum^{n}_{i=1}z_i + \bar{\gamma} -( \bar{\gamma} - \bar{\alpha})(\bar{\gamma}-\bar{\alpha}-1)+\sigma_1\right)2\hbar^2 k,\label{enr-nlo}
\end{eqnarray}
with 
\begin{equation}
\sigma_1 = \mu \sum^{n}_{i} z^2_i + (2- \mu)\sum^{n}_{i = 1} z_i-2n^2 -2s-\frac{l(l+1)}{4}+(\bar{\gamma}-\bar{\alpha})\left(\bar{\gamma}-\bar{\alpha}-\frac{3}{2}\right). 
\label{sigman1_r}
\end{equation}
Since for $\displaystyle{s = -\frac{l}{2}}$ we get unbounded solutions, we consider $\displaystyle{s = \frac{l+1}{2}}$  and eigenfunctions for the system (\ref{ham-nlo}) are 
\begin{equation}
\Phi_{n,l}(r,\theta, \phi) = {\cal N}_{n,l}\;e^{\frac{-\omega_0 r^2}{2\;\hbar(1 + k r^2)}}(k r^2)^{\frac{l}{2}}(1 + k r^2)^{n + \bar{\gamma}-\bar{\alpha}} \Pi^{n}_{i=0}\left(\frac{k\;r^2}{(1 + k r^2)} - z_i\right),\label{phirnl}
\end{equation}
where ${\cal N}_{n,l}$ is the normalization constant and the roots can be found out by using the relation, 
\begin{equation}
\sum^n_{j\neq i} \frac{2}{z_i - z_j}= \frac{\mu z_i^2 + (2n-\mu) z_i + \left(l + \frac{3}{2}\right)}{z_i(z_i - 1)}.\label{bethe-ansatz-nlo2}
\end{equation}
The energy eigenvalues (\ref{enr-nlo}) are also of quadratic nature similar to the Higgs oscillator and Mathews-Lakshmanan oscillator. While solving the system (\ref{schwingerq-nlor}) quantum mechanically the ordering parameters play an important role since $\sigma_1$ and $\sigma_2$ (vide (\ref{sigma1}) and (\ref{sigma2})) are also being related with $n$. This is possible when the ordering parameters are treated as arbitrary. To obtain the eigenfunctions for different values of $n$, we have to find out the roots $z_i, \; i = 0, 1, 2, ...n$,\ by solving the equation in $z_i$ of degree $n+1$. Hence it is possible only to obtain a few states of the system (\ref{ham-nlo}) which have been listed in the Table given in Appendix B.  
  
As in the case of one dimensional system (\ref{abeta34}), we again fixed the ordering parameters to normalize the solutions (\ref{phirnl}) for different combinations of $n$ and $l$.  
For $n = 2$, we will get the coupled equations of $z_1$ and $z_2$ from (\ref{sigman1_r}) and (\ref{bethe-ansatz-nlo2}) which can not be solved analytically. Hence, the system (\ref{ham-nlo1}) is quasi exactly solvable. 

\section{\label{conc}Conclusion}
We have studied the classical and quantum dynamics of two one dimensional nonlinear oscillators, namely the (i) Higgs oscillator  and (ii) a $k$-dependent nonpolynomial rational potential. Both the systems share the same mass functions, but differ in their potential forms and the dynamics of the two systems are different. We obtained explicitly the solutions of the Higgs oscillator and the non-polynomial $k$-dependent potential for their classical dynamics. The former solution is expressed in terms of  trigonometric functions, whereas for the second system the classical solution can be given in terms of Jacobian elliptic functions. The $k$-dependent nature of the potential gives rise to a complex solution. In the quantum study, the Higgs oscillator is exactly solvable for all possible choices of ordering parameters, whereas the second system is quasi exactly solvable  and it possesses quasi-polynomial solutions with a requirement that the ordering parameters are arbitrary.  We also studied the classical and quantum dynamics of the three dimensional generalization of the above two nonlinear systems. 

\newpage
\begin{figure}[h]
    \vspace*{-2cm}
    \makebox[\linewidth]{
        \includegraphics[width=1\linewidth]{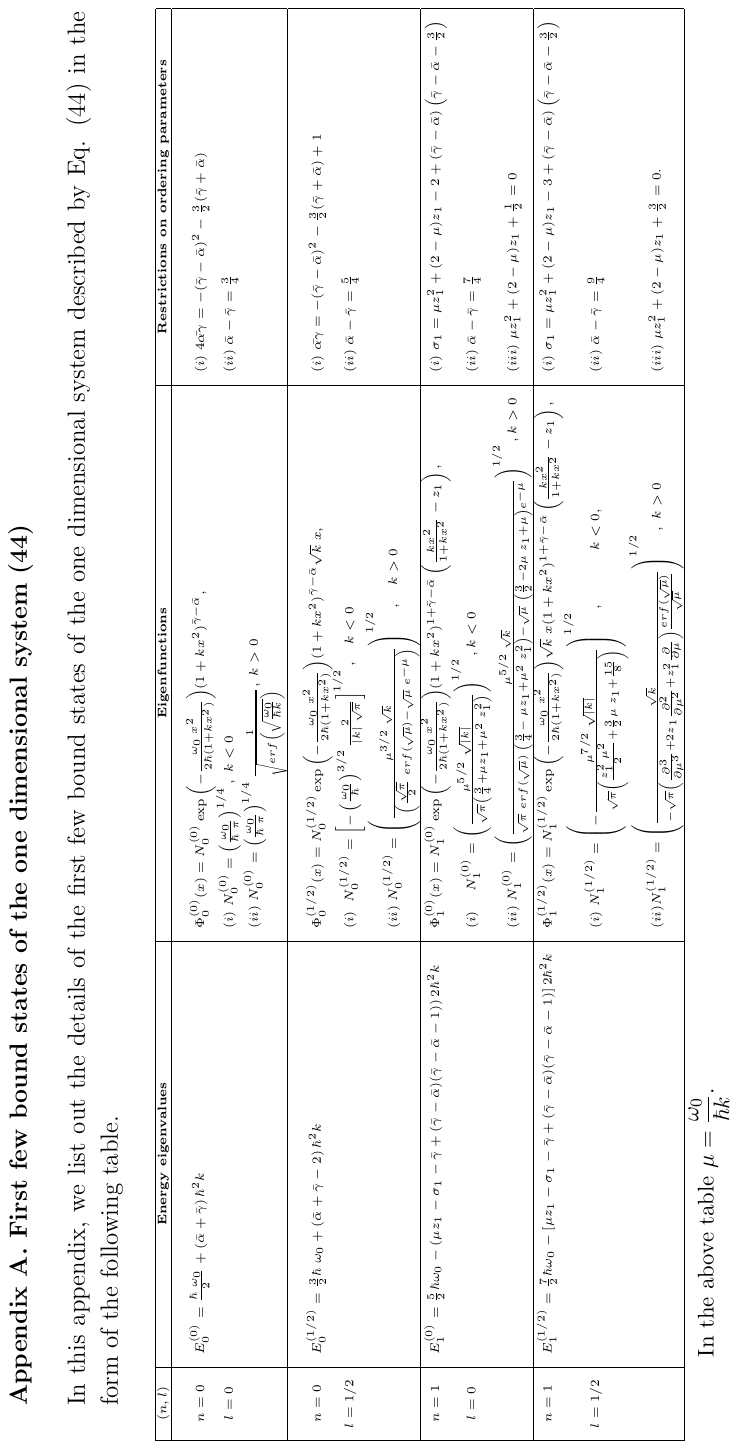}
    }    
\end{figure}

\newpage
\begin{figure}[h]
    \vspace*{-2cm}
    \makebox[\linewidth]{
        \includegraphics[width=1\linewidth]{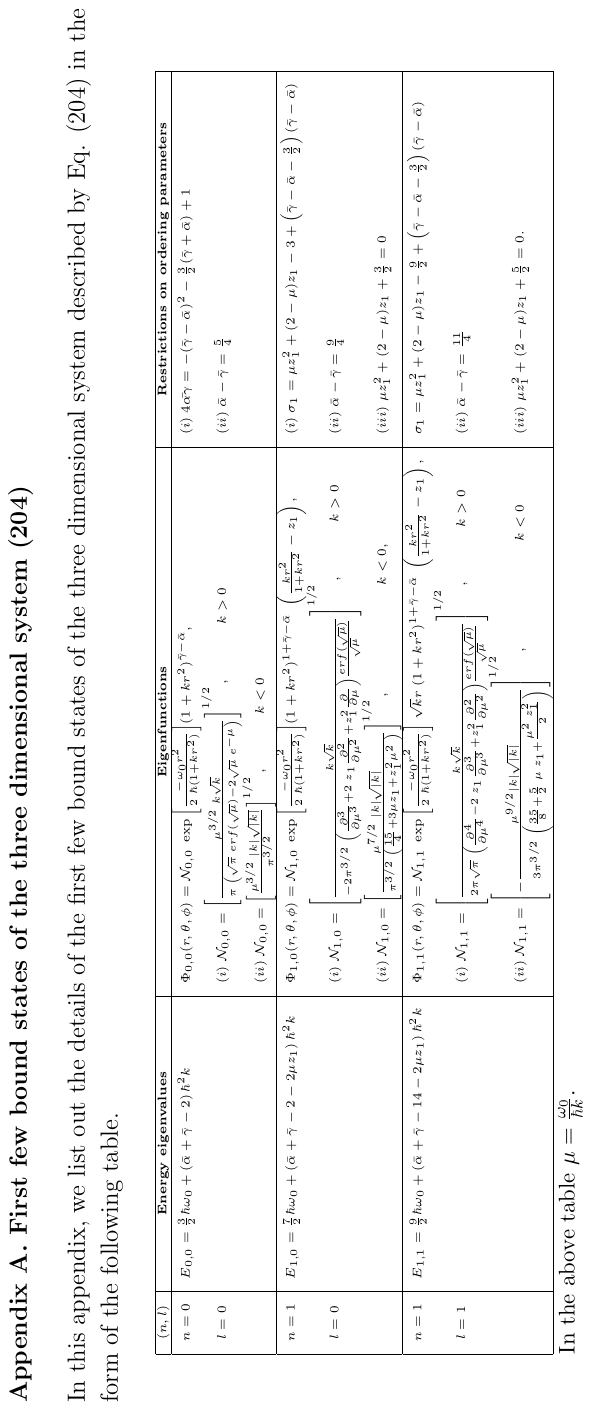}
    }
    
\end{figure}

\section*{Acknowledgement}
VC wishes to acknowledge DST for the financial support of the project (No. SR/WOS-A/PM-64/2018(G)) under Women Scientist Scheme A. ML acknowledges the financial support under a DST-SERB Distinguished Fellowship program (Grant No. SB/DF/04/2017).

\end{document}